\documentclass[prl,showpacs,preprintnumbers,amsmath,amssymb,twocolumn,superscriptaddress,notitlepage]{revtex4-2}

\usepackage[utf8]{inputenc}

\usepackage[normalem]{ulem}
\usepackage {marvosym}
\usepackage{ifsym}

\usepackage{graphicx}
\usepackage[usenames,dvipsnames]{color}
\usepackage[
colorlinks=true,       
linkcolor=red,          
citecolor=blue,        
filecolor=magenta,      
urlcolor=cyan  
]{hyperref}

\usepackage{dcolumn}
\usepackage{bm}
\usepackage{endnotes}

\hypersetup{colorlinks,citecolor= blue,linkcolor= blue, urlcolor=blue}

\usepackage{amsmath}
\usepackage{bbm}
\usepackage{amsfonts}
\usepackage{amssymb}
\usepackage{latexsym}
\usepackage{graphicx}
\usepackage[english]{babel}
\usepackage{multirow}
\usepackage{float}
\usepackage{url}
\usepackage{slashed}
\usepackage{xcolor} 
\usepackage{array}
\usepackage{cases}

\newcommand{\be}{\begin{equation}}
\newcommand{\ee}{\end{equation}}
\newcommand{\ba}{\begin{array}}
\newcommand{\ea}{\end{array}}
\newcommand{\bea}{\begin{eqnarray}}
\newcommand{\eea}{\end{eqnarray}}

\usepackage{ulem,fancyvrb}
\usepackage{xcolor}

\begin{document}

\title{
Amplifying nonresonant production of dark sector particles \\
in scattering dominance regime
}

\author{Mingxuan Du} 
\thanks{Corresponding author}
\email{mingxuandu@pku.edu.cn}
\affiliation{School of Physics and State Key Laboratory of Nuclear Physics and Technology, Peking University, Beijing 100871, China}
\affiliation{Center for High Energy Physics, Peking University, Beijing 100871, China}

\author{Jia Liu}
\thanks{Corresponding author}
\email{ jialiu@pku.edu.cn}
\affiliation{School of Physics and State Key Laboratory of Nuclear Physics and Technology, Peking University, Beijing 100871, China}
\affiliation{Center for High Energy Physics, Peking University, Beijing 100871, China}

\author{Xiao-Ping Wang}
\thanks{Corresponding author}
\email{hcwangxiaoping@buaa.edu.cn}
\affiliation{School of Physics, Beihang University, Beijing 100083, China}
\affiliation{Beijing Key Laboratory of Advanced Nuclear Materials and Physics, Beihang University,
Beijing 100191, China}

\author{Tianhao Wu}
\thanks{Corresponding author}
\email{tianhaowu@stu.pku.edu.cn}
\affiliation{School of Physics and State Key Laboratory of Nuclear Physics and Technology, Peking University, Beijing 100871, China}

\begin{abstract}
We investigate the enhancement of dark sector particle production within the scattering dominant regime. These particles typically exhibit a slight mixing with Standard Model particles through various portals, allowing for their generation through in-medium oscillation from Standard Model particle sources. Our analysis reveals that in the scattering dominance regime, with a significantly smaller scattering mean free path $\lambda_{\rm sca}$ compared to the absorption mean free path $\lambda_{\rm abs}$, the nonresonant production of sterile states can experience an enhancement by a factor of $\lambda_{\rm abs}/\lambda_{\rm sca}$. This phenomenon is demonstrated within the context of kinetic mixing dark photon production at a reactor, precisely satisfying this condition.
By incorporating this collisional enhancement, we find that the current sensitivity to the mixing parameter $\epsilon$ for dark photons in the TEXONO experiment can be significantly improved across a range spanning from tens of eV to MeV. This advancement establishes the most stringent laboratory constraint within this mass spectrum for the dark photon.
Sterile neutrino production, however, does not exhibit such enhancement, either due to the failure to meet the scattering dominance criterion or the neutrino damping in resonant production.
\end{abstract}

\maketitle

\section{INTRODUCTION}

The dark photon $A'$ is a well-motivated new physics particle, which is the gauge boson corresponding to an extra $U(1)'$ gauge symmetry and can couple to the Standard Model (SM) through a kinetic mixing term~\cite{FAYET1980285, Fayet:1980rr, Okun:1982xi, Galison:1983pa, Holdom:1985ag, Fayet:1990wx, Foot:1991kb}. After normalization of kinetic terms, it couples to the electric charge of the SM particle with a mixing strength $\epsilon$, serving as a bridge between the SM and the dark sector~\cite{Pospelov:2007mp, Arkani-Hamed:2008hhe, Essig:2013lka, Alexander:2016aln, Battaglieri:2017aum} or a dark matter candidate as well~\cite{Redondo:2008ec, Nelson:2011sf, Arias:2012az, Graham:2015rva}. Its mass can be originated from the Stueckelberg or the Higgs mechanisms~\cite{Kors:2005uz, Feldman:2006ce, Feldman:2006wb, Feldman:2007wj, Feldman:2009wv, Du:2019mlc, Du:2021cmt, Redi:2022zkt}, thus there are only two phenomenology parameters, mass $m_{A'}$ and interaction strength $\epsilon$ in the model. Recent experiments have extensively studied the parameter space from particle physics, astrophysics and cosmology, see Refs.~\cite{Fabbrichesi:2020wbt, Caputo:2021eaa} for a recent review on the dark photon and the various constraints.

Among various terrestrial laboratories~\cite{Demidov:2018odn, Ilten:2018crw, Bauer:2018onh}, nuclear reactor experiments such as TEXONO~\cite{Danilov:2018bks} and JUNO-TAO~\cite{Smirnov:2021wgi}, can provide the competitive limits on the dark photon in the mass range of ($10$ eV $\sim$ MeV). At the core of a nuclear reactor, numerous photons with MeV energy are produced, which can potentially transform into dark photons with masses up to MeV if allowed by the kinematics. 
If the dark photon mass is well below the effective photon mass $m_\gamma$ (which is around 10 eV in the reactor~\cite{Danilov:2018bks}), then the conversion probability will suffer the suppression $\sim m_{A'}^4/m_\gamma^4$~\cite{Mizumoto:2013jy, Fortin:2019npr}. Consequently, the nuclear reactor experiments are 
the suitable place to explore the high mass range $10$ eV $\sim$ MeV.

Estimating dark photon production in nuclear reactor cores is performed in the interaction basis, where the dark photon is sterile to the SM electromagnetic current. Consequently, the dark photon can be generated only through the oscillation between the SM photon and the dark photon when photons are produced in the reactor. This scenario requires considering the interaction between the SM photon and reactor cores, which directly affects the oscillation process. For photons in reactor cores, two important effects are addressed: absorption, which terminates the oscillation, and scattering with the medium, which destroys the old photon and generate a new one with a softer energy. We emphasize that the scattering rate is significantly larger than the absorption rate for MeV reactor photons \cite{ParticleDataGroup:2022pth}, resulting in multiple scatterings before absorption occurs. Previous studies only considered absorption when calculating the conversion of photons to dark photons in the reactor \cite{Danilov:2018bks, Seo:2020dtx}. However, this study will also include the scattering effects and examine how they modify the results.
\\

\section{The dark photon and photon oscillation scheme}

In the kinetic-mixing dark photon model \cite{Holdom:1985ag}, the Hamiltonian for the transverse modes of the photon and dark photon in the medium 
in the interaction basis
is given by \cite{Redondo:2013lna}
\begin{eqnarray}
H=\frac{1}{2 E_\gamma}\left(\begin{array}{cc}
\epsilon^2 m_{A'}^2+m_\gamma^2 & -\epsilon m_{A'}^2 \\
-\epsilon m_{A'}^2 & m_{A'}^2
\end{array}\right), 
\label{eq: int Ham}
\end{eqnarray}
where the dark photon $A'$ has mass $m_{A'}$,
$E_\gamma$ is the (dark) photon energy,
and $m_\gamma=\sqrt{4 \pi \alpha n_e / m_e}$ is the effective mass of photon induced by the medium, which is about $m_\gamma = 20$ eV \cite{Danilov:2018bks} in the reactor core. Our analysis focuses only on the non-resonance region, specifically when $m_{A'} \gg m_\gamma$ \cite{Danilov:2018bks}, as exploring the resonance region requires a detailed understanding of the reactor core structure and materials.

We do not include the longitudinal photon in our calculation because its energy, equal to the plasma frequency (approximately 20 eV), is below the threshold of the neutrino detector used in this study. Additionally, the longitudinal plasmon is not a propagation mode, so the physical picture of Compton scattering during propagation is not valid. Therefore, we only consider the transverse mode of the photon in our study.
\\

\section{The absorption and scattering in the medium}

We will examine matter effects of photon oscillations to dark photons in a medium, including absorption and scattering.  
When dealing with MeV photons in the reactor, we will consider the photoelectric process for absorption and Compton scattering for scattering as they are the dominant processes~\cite{ParticleDataGroup:2022pth}.

The damping processes (absorption and scattering) terminate photon propagation and oscillation to the dark photon. 
The conversion probability with damping effect is given by \cite{Redondo:2015iea}
\begin{equation}
P_{\gamma \to A'}(\ell, \Gamma
, E_\gamma) 
= \epsilon^2 m_{A'}^4 \frac{1+e^{-\Gamma \ell}-2 \cos \frac{\Delta m^2 \ell}{2 E_\gamma} e^{-\frac{\Gamma \ell}{2}}}{\left(\Delta m^2\right)^2+(E_\gamma \Gamma)^2},
\label{eq: ab prob}
\end{equation}
where the damping rate is $\Gamma = \Gamma_{\rm abs} + \Gamma_{\rm sca}$.
Here
$\Gamma_{\rm abs} = n_A \sigma_{{pe}}$ represents the absorption rate for the photoelectric process,
and $\Gamma_{\rm sca} = n_A \sigma_C$ is the Compton scattering rate.
Assuming a pure thorium ($Z=90$) reactor core \cite{Dent:2019ueq},
$n_A$ is the corresponding number density of thorium atoms, and $\sigma_{pe}$ and $\sigma_C$ denote the cross section for photoelectric process and Compton scattering in pure thorium \cite{Berger:xcom}. 
By neglecting the scattering rate and consider a constant absorption rate $\Gamma_{\rm abs} \simeq 1/(10 \rm{cm})$ for MeV photons in reactor cores \cite{Danilov:2018bks}, for $\ell \to \infty$, Eq.~\eqref{eq: ab prob} recovers to \cite{Redondo:2008aa, An:2013yfc, Redondo:2013lna, Redondo:2015iea}
\begin{equation}
P_{\gamma \to A'}^{\rm{a }} = \epsilon^2 \frac{ m_{A'}^4}{(\Delta m^2)^2 + E_\gamma^2 \Gamma_{\rm abs}^2},
\label{eq: prob limit}
\end{equation}
which is the conversion probability of a photon oscillating to a dark photon at infinity solely under the influence of absorption.

In the oscillation calculation, we adopt a hard cutoff at the kinetic threshold in line with Ref.~\cite{Danilov:2018bks}. This approach adheres to the forward scattering of the Compton-like conversion process, $\gamma + e \to A' + e$, where the energy and momentum direction of the incoming photons and outgoing dark photons are assumed to be identical. Consequently, the kinetic threshold for the conversion process is defined as $E_{\gamma} > m_{A'}$, ensuring that the oscillation probability only considers cases where the incoming photon has sufficient energy to convert into a dark photon.

Note that $\Gamma_{\rm sca} \gg \Gamma_{\rm abs}$ for MeV photons in reactor cores \cite{ParticleDataGroup:2022pth, Berger:xcom}. Therefore, the scattering process $\gamma + e \to \gamma + e$ will significantly impact the oscillation by absorbing and emitting photons, altering their energy and direction. It is essential to incorporate these newly released photons into the simulation and recalculate the oscillation probability.
The reactor core produces isotropic photons \cite{Bechteler:1984}, and we assume a homogeneous core \cite{Dent:2019ueq} that is much larger [$O(\rm{m})$] than the absorption length [$O(10~\rm{cm})$] of MeV photons \cite{Danilov:2018bks}.
These assumptions lead us to expect isotropic distribution of resulting reactor dark photons. Thus, our analysis focuses solely on the resulting dark photon spectrum. Notably, a photon becoming a dark photon retains its original energy during oscillation. Therefore, our study considers only energy changes in Compton scattering as the direction change does not affect the dark photon energy spectrum.
\\

\section{Simulation process}

In our simulation, the absorption and scattering effects in the reactor cores are explored by tracking the individual photon. 
We consider a photon generated from the reactor core with initial energy $E_1$. 
This photon first travels through a mean free path of a Compton scattering process, $\lambda_C $, in pure thorium medium
\cite{Berger:xcom}
\begin{equation}
\lambda_C (E_1) = 1 / (n_A \sigma_C ( E_1)).
\end{equation}
For a propagation distance smaller than $\lambda_C $, no Compton scattering is assumed to take place.
Then the probability of the photon oscillating to a dark photon with absorption effect in this distance is given by $P_{\gamma \to A',\ 1}(\lambda_C (E_1), \Gamma_{\rm abs} (E_1), E_1)$ as Eq.~\eqref{eq: ab prob}, where the subscript $1$ records the step of the propagation. 
Considering loss of photon due to the medium absorption and the oscillation to dark photon, the resulting survival probability for photon is 
\begin{eqnarray}
P^s_1 =
\left[1 - P_{\gamma \to A',\ 1} \right] e^{-\lambda_C \Gamma_{\rm abs} }.
\end{eqnarray}

After traveling the mean free path distance $\lambda_C$, the photon is assumed to undergo next Compton scattering immediately, where the old photon with energy $E_1$ is absorbed by the medium and a new photon with lower energy $E_2 < E_1$ is generated. The energy $E_2$ of the new photon is randomly chosen based on the final state photon energy distribution in the laboratory frame for Compton scattering.

\begin{figure}[htbp]
\begin{centering} 
\includegraphics[width=1 \columnwidth]{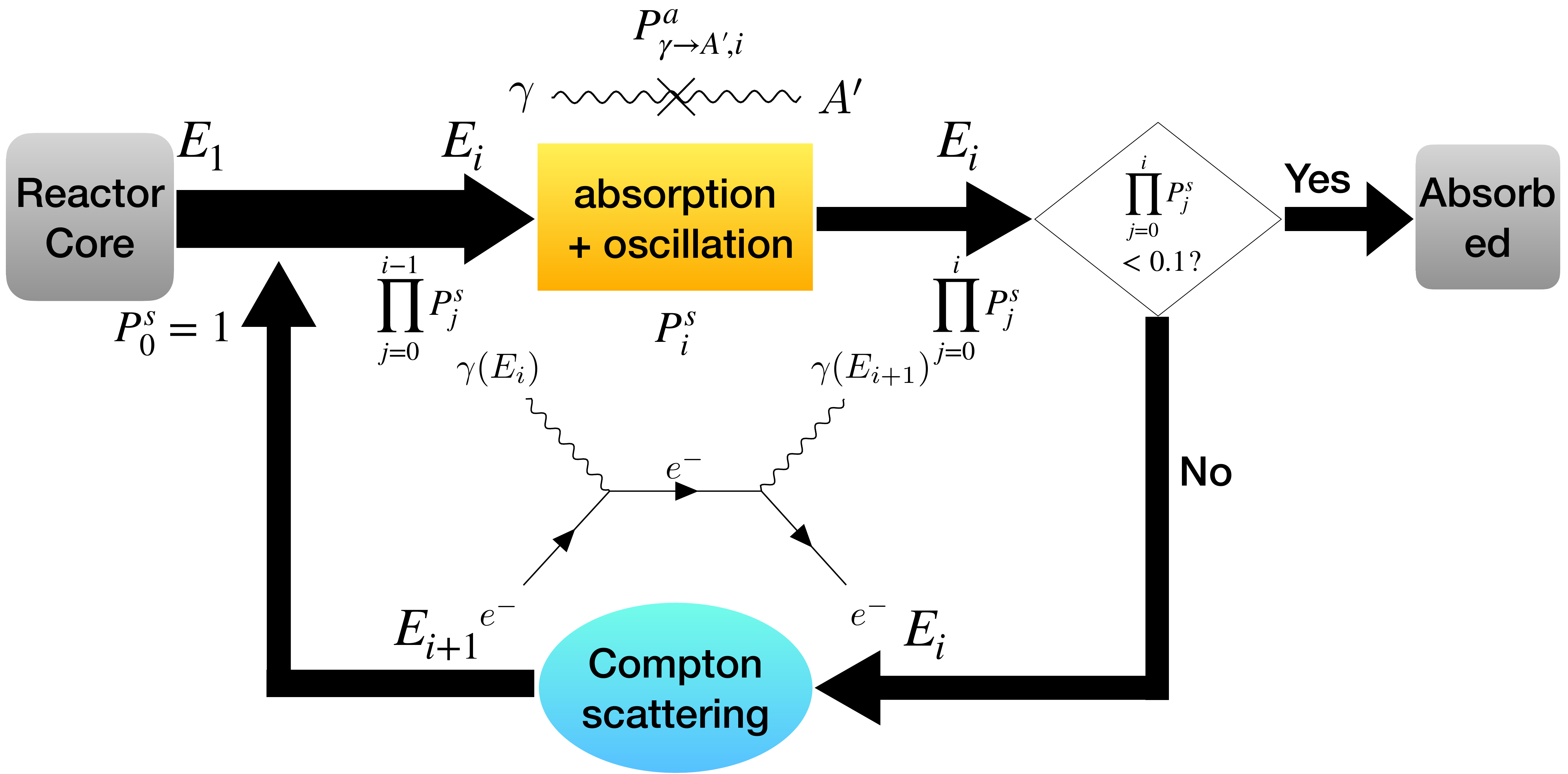} 
\caption{The flow chart of our simulations for 
reactor photons oscillating to dark photons 
with Compton scattering included.
}
\label{fig: Flow Chart}
\end{centering}
\end{figure}

In the next step, the new photon travels through the medium over another mean free path distance $\lambda_C(E_2)$ and undergoes the processes of propagation, absorption, oscillation, and scattering, see Fig.~\ref{fig: Flow Chart} 
for our simulation flow chart.
We loop these processes until the survival probability of the final photon falls below 0.1, denoted as $\Pi_{i=0}^n P^s_i$ (with initial value $P^s_0 = 1$), 
where $i$ is the step index 
and $n$ is the number of loops when the survival probability falls below 0.1.  
Here we choose 0.1 as the threshold to reduce the computing power requirement, 
since a smaller number will not significantly change the production of dark photons.

\begin{figure}[htbp]
\begin{centering} 
\includegraphics[width=0.98 \columnwidth]{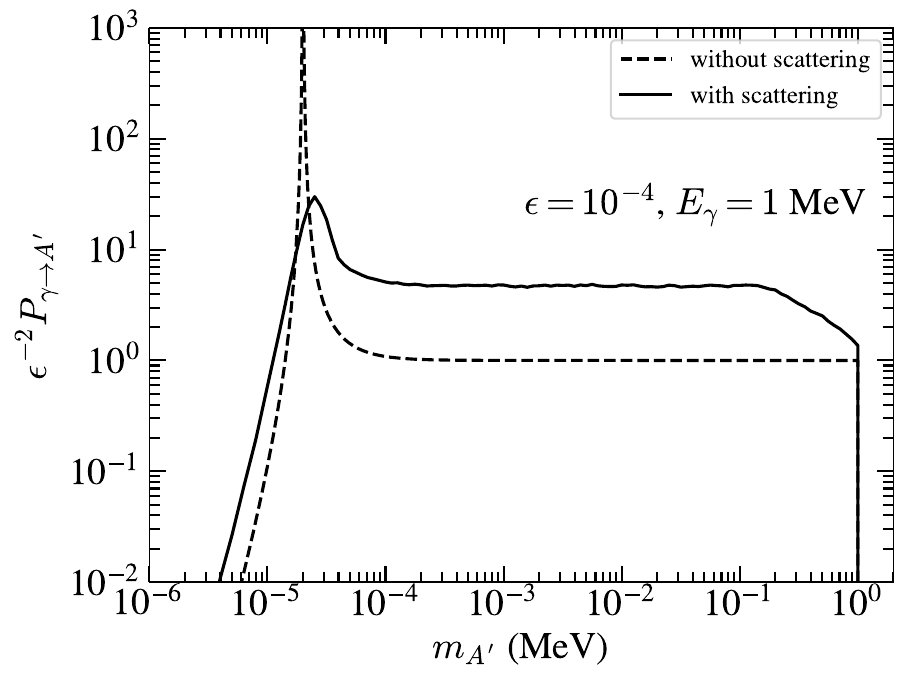}
\caption{The rescaled conversion probability for a photon with energy 1 MeV to dark photon in reactor cores under different scenario. The solid curve considers both the absorption and Compton scattering effects. The dashed curve only considers the absorption effect. 
}
\label{fig:probability}
\end{centering}
\end{figure}

Finally, after considering absorption and scattering effects, we simulated the total conversion probability from a photon to a dark photon in the medium, which can be expressed as
\begin{equation}
P_{\gamma \to A'} = \sum_{i=1}^n \left(P^a_{\gamma \to A',\ i} \times \Pi_{j=0}^{i-1} P^s_j \right).
\label{eq: prob single photon}
\end{equation}
It sums up the probabilities of generating dark photons in each step along the photon propagation path.
Furthermore, to suppress the fluctuations in our results,
we simulated $1000$ photons for each initial energy. In Fig.~\ref{fig:probability}, when counting the scattering effect, the conversion probability is significantly enhanced for $m_{A'} > 20$ eV. 
\\

\section{Analytic estimation}

To comprehend the simulation results, we aim to derive an analytical explanation for the observed enhancement in the total conversion probability.
Equation~\eqref{eq: ab prob} contains two lengths for the MeV photon: 
interaction length $\lambda = 1/\Gamma \sim 2$ cm 
and oscillation length $ \ell_{\rm osc} = 4 \pi E_\gamma / \Delta m^2 \sim 10^{-10}(m_{A'}/ \rm{MeV})^{-2}$ cm. 
Therefore, $\ell_{\rm osc}$ leads to a highly oscillating pattern with propagation length $\ell$. 
To simplify, we disregard the highly oscillating linear term $\cos{(l/l_{\rm osc})}$ in Eq.~\eqref{eq: ab prob}, 
and obtain an average estimation of the single conversion probability. 
Then the average probability 
of a photon not undergoing any scattering and converting to a  nonresonant dark photon is
\begin{equation}
P_0(\ell)
= \frac{\epsilon^2 m_{A'}^4\left(1+e^{-\Gamma \ell}\right)}{\left(\Delta m^2\right)^2+(E_\gamma \Gamma)^2} 
\simeq \frac{\epsilon^2 m_{A'}^4\left(1+e^{-\Gamma \ell}\right)}{\left(\Delta m^2\right)^2}.
\label{eq: 0 sca P}
\end{equation}
We drop the $E_\gamma \Gamma$ term in the denominator since
$\Delta m^2 \sim m_{A'}^2 \gg E_\gamma \Gamma \sim (3 ~\rm{eV})^2$ 
for nonresonant dark photon production at the reactor,
resulting in an energy-independent conversion probability.

For a rough estimate,  
assuming no energy loss during medium scattering,
the probability 
of a photon undergoing $n$ times of scatterings
at locations $0 < \ell_1 < \ell_2, ... < \ell_n < \ell$
and converting to a nonresonant dark photon is
({see a detailed derivation in the Appendix \ref{appsec: prob-n derivation}})
\begin{eqnarray}
P_n (\ell) &=& \int_{0}^\ell d \ell_1 
\int_{\ell_1}^\ell d \ell_2 ... 
\int_{\ell_{n-1}}^\ell d \ell_n
e^{-\Gamma \ell_n} \Gamma_{\rm sca}^n P_0(\ell -\ell_n) \nonumber \\
&=& \frac{\epsilon^2 m_{A'}^4}{(\Delta m^2)^2} \frac{\Gamma_{\rm sca}^n}{\Gamma^n}
\left(1 +e^{-x} \frac{x^n}{n!} - \frac{\Gamma(n,x)}{\Gamma(n)} \right),
\label{eq: prob-n-sca}
\end{eqnarray}
where $x=\Gamma \ell$,
while $\Gamma(n,x)$ and $\Gamma(n)$ are the incomplete and complete Gamma functions, respectively. 
The damping effect on the active state at the $n$th scattering is described by $e^{-\Gamma \ell_n}$,
$\Gamma_{\rm sca}^n$ represents the $n$ times of scatterings,
and the last term $P_0(\ell -\ell_n)$ gives the conversion probability
from the new photon generated by the $n$th scattering. 

The total conversion probability,
summing up the multiscattering between photons and medium, leads to
\begin{eqnarray}
P^{\rm tot} &=& \sum_{n=0}^{\infty} P_n 
= \frac{\epsilon^2 m_{A'}^4}{(\Delta m^2)^2} 
\left(\frac{\Gamma}{\Gamma_{\rm abs}} 
+ B(\ell)\right),  \\
B(\ell) &\equiv&  e^{-\Gamma_{\rm abs} \ell} - \sum_{n=1}^{\infty} \frac{\Gamma_{\rm sca}^n}{\Gamma^n}\frac{\Gamma(n,\Gamma \ell)}{\Gamma(n)}. \nonumber
\end{eqnarray}
For $\ell \to 0$, $P^{\rm tot} (0) =  P_0 (0)$ thus the short propagation distance suppresses the multiple scattering effect.
When the active state propagates long distance in the medium, $\ell~ (\mathcal{O}(\text{m})) \gtrsim  1/\Gamma_{\rm abs} ~(\mathcal{O}(10 ~ \text{cm})) $,
one has 
\begin{equation}
P^{\rm tot} (\ell) \approx \frac{\Gamma}{\Gamma_{\rm abs}}  P_0 (\ell \to \infty).
\end{equation}
In the scattering dominant region
where $\Gamma \sim \Gamma_{\rm sca} \gg \Gamma_{\rm abs}$,
the nonresonant dark photon production
is enhanced by the factor
$\Gamma_{\rm sca} / \Gamma_{\rm abs} \sim \lambda_{\rm abs} / \lambda_{\rm sca}$
compared to the case without considering scattering.
For MeV photons in reactor, this enhancement factor is about $10~\rm{cm} / 2~\rm{cm} \sim 5$, which in the ballpark agrees with 
the factor of $10$ in our full numeric simulation for the nonresonant region $m_{A'} > 50$ eV in Fig.~\ref{fig:probability}. 

 In Fig.~\ref{fig:probability}, the solid line representing the conversion probability exhibits a slight decrease near the kinetic threshold, primarily attributed to the energy loss of photons during elastic Compton scattering. In contrast, the dashed line, which excludes scattering effects, maintains a constant value. Moreover, the asymptotic value of the dashed line approaches unity, indicating that the converted dark photon flux in the absence of scattering scales directly with the initial photon flux, $\Phi_{A'} \approx \epsilon^2 \Phi_{\gamma}$. When incorporating scattering effects, the dark photon flux is enhanced by a factor of $\Gamma/\Gamma_{\rm abs} \sim 5$, highlighting the significance of scattering in determining the overall dark photon flux in a reactor core environment.

Comparing $\gamma - A'$ mixing with sterile neutrino mixing is intriguing. Unlike nonresonant dark photons in reactors, the production of sterile neutrinos in supernovae lacks a similar enhancement. Two mechanisms drive sterile neutrino production in supernovae: adiabatic MSW resonance and collisional production \cite{PhysRevD.36.2273, PhysRevD.83.093014, PhysRevD.99.043012}. The dominant mechanism, adiabatic MSW resonance, loses efficacy due to a high scattering rate, significantly reducing neutrino mean free path~\cite{PhysRevD.99.043012}. Conversely, collisional production is less influenced by scattering rate, as the damping factor of $\Gamma^{2}$ in conversion probability is counterbalanced by scattering rate in the numerator and enhanced resonant length~\cite{Redondo:2008aa}. Thus, elevated scattering rates generally suppress sterile neutrino production in supernovae. In contrast, for reactor dark photons in the nonresonant regime, the mass term $(\Delta m^2)^2$ outweighs the significance of the scattering damping term $E^2\Gamma^{2}$ in the conversion probability. Therefore, increased scattering rates linearly amplify dark photon production at the reactor.
\\

\section{The reactor experiment search of dark photons}

\begin{figure}[htbp]
\begin{centering} 
\includegraphics[width=0.98 \columnwidth]{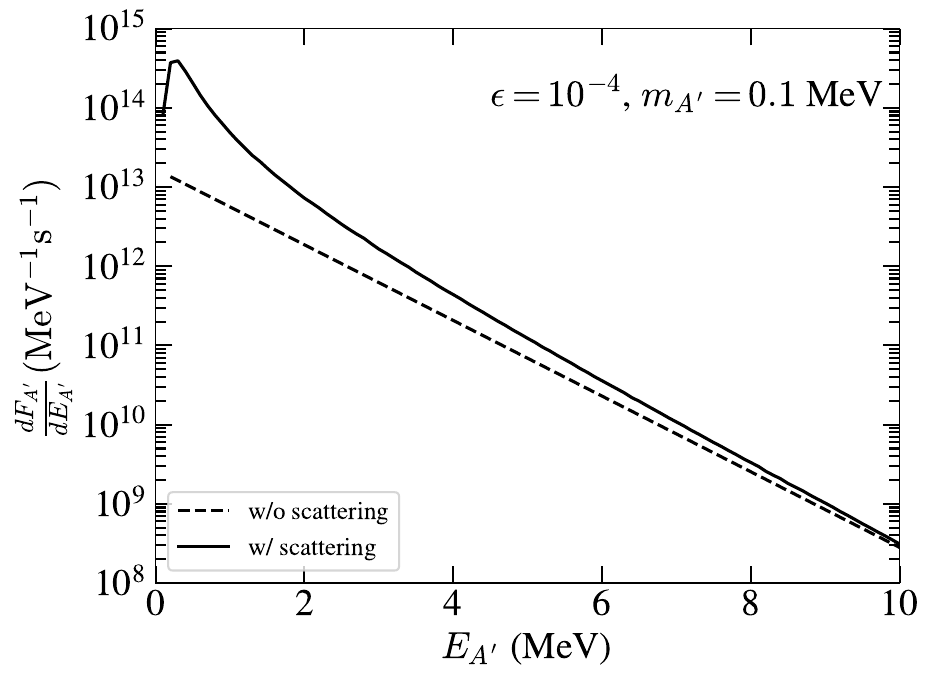}
\caption{The spectra of dark photon with $m_{A'} = 0.1$ MeV produced in the reactor with 2.9 GW thermal power under different scenarios.
The solid curve considers both the absorption and
Compton scattering.
The dashed curve only considers
the absorption effect.
}
\label{fig: DP production rate}
\end{centering}
\end{figure}

Nuclear reactor cores produce numerous MeV photons, making them ideal for dark photon production. Assuming isotropic generation of prompt photons, the approximate spectrum for photon flux at $E_\gamma > 0.2$ MeV is
\cite{Bechteler:1984}
\begin{equation}
\frac{dF_\gamma}{dE_\gamma} = \frac{0.58 \times 10^{21}}{\mathrm{s}\, \mathrm{MeV}}\left(\frac{P}{\mathrm{GW}}\right) \exp\left[-\frac{1.1E_\gamma}{\mathrm{MeV}}\right],
\label{eq: gamma flux}
\end{equation}
where $P$ represents the reactor thermal power.
The dark photon production rate from photon oscillation can be expressed as
\begin{equation}
\frac{dF_{A'}}{dE_{A'}} = \int dE_\gamma \frac{dF_\gamma}{dE_\gamma} \frac{dP_{\gamma \to A'}}{dE_{A'}}(E_\gamma),
\label{eq: A' spectra}
\end{equation}
where ${dP_{\gamma \to A'}}/{dE_{A'}}$ represents the differential conversion probability in our simulation. 
Figure~\ref{fig: DP production rate} illustrates the energy spectrum for dark photons produced by a 2.9 GW nuclear reactor with $m_{A'} = 0.1$ MeV. 
The solid line accounts for absorption and scattering effects,
while the dashed line only considers absorption.
Since the scattering softens the photon energy and $E_\gamma = E_{A'}$ in the oscillation,
more low-energy $A'$ are produced with inclusion of scatterings, 
resulting in a larger enhancement for the lower energy dark photon in the solid line compared to the dashed line. 

\begin{figure}[htbp]
\begin{centering} 
\includegraphics[width=0.98 \columnwidth]{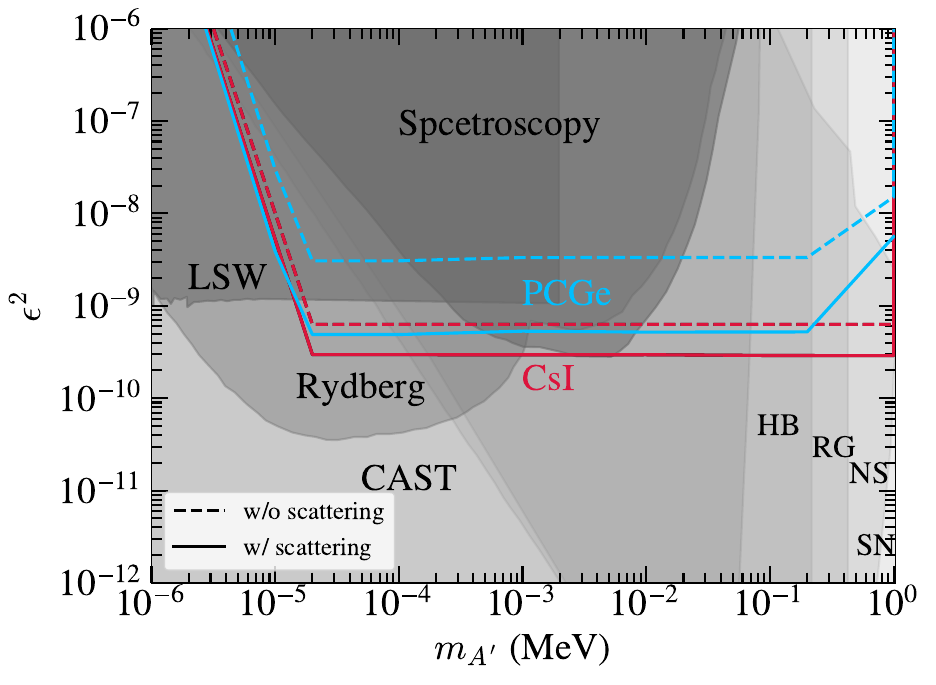}
\caption{TEXONO $95\%$ C.L. constraints on dark photons lighter than $1$ MeV from CsI (red) and PCGe (blue) detectors.
The solid curves are the constraints incorporating both absorption and scattering effects on photons by the medium, and the dashed curves incorporate only the absorption. Existing constraints from terrestrial experiments are shown as dark gray regions 
\cite{Jaeckel:2010xx, INADA2013301}.
Astrophysical and cosmological constraints are shown as light gray regions \cite{Redondo:2008aa, Redondo:2013lna, Vinyoles:2015aba, Chang:2016ntp, Hong:2020bxo}. 
}
\label{fig:constraint}
\end{centering}
\end{figure}

The dark photons, with negligible attenuation due to oscillation back into photons ($\epsilon^2 \ll 1$), can be detected through Compton scattering of electrons once they reach the detector
by measuring the electron recoil energy.
We analyzed experimental data \cite{TEXONO:2009knm, TEXONO:2014eky, TEXONO:2018nir} from high-energy threshold ($3 \sim 8$ MeV) CsI(Tl) scintillating detectors \cite{TEXONO:2009knm} and low-energy threshold ($0.3 \sim 12.3$ keV) Germanium (Ge) ionization detectors \cite{TEXONO:2014eky} at TEXONO \cite{TEXONO:2000zzq}.
Then in Fig.~\ref{fig:constraint},
we set $95\%$ C.L. {(confidential level)} constraints on dark photons with a mass lighter than $1$ MeV obtained from the CsI (red) and PCGe (blue) detectors in the TEXONO experiment, {see the Appendix \ref{appsec: reactor search} for more details}. For dark photon heavier than $1$ MeV, it can quickly decay to a pair of electrons thus does not apply to the scenario described here.
The solid curves represent the constraints considering both absorption and scattering effects of the medium on photons, while the dashed curves only consider the absorption effect. 
\\

\section{Results and discussion}

The inclusion of scatterings increases the conversion probability of oscillation into the dark photon by a factor of 10 in Fig.~\ref{fig:probability}. 
The number of signals $N_{\text{sig}}$ in the detector is proportional to $\epsilon^4$, 
so the $\epsilon^2$ limit will be enhanced about $\sqrt{10} \sim 3$ times. 
Additionally, scattering softens the energy of the photons, 
reducing $N_{\text{sig}}$ for high-energy threshold detectors (e.g., CsI detector with threshold of $3 \sim 8$ MeV) 
and increasing $N_{\text{sig}}$ for low-energy threshold detectors (e.g., PCGe detector with threshold of $0.3 \sim 12.3$ keV). 
Consequently, the $\epsilon^2$ limit is enhanced by approximately two times for CsI detectors limits, and si times for PCGe detectors limits. This highlights the significance of considering the scattering effect in dark photon constraints. Consequently, TEXONO provides the most stringent constraints on dark photons among terrestrial experiments \cite{Jaeckel:2010xx, Schwarz:2015lqa, Fabbrichesi:2020wbt} 
Within the range of dark photon masses spanning from 10 eV to MeV. 

The photon energy spectrum in Eq.~\eqref{eq: gamma flux} only includes photons with energies $E_\gamma>0.2$ MeV. 
Thus, when $m_{A'}>0.2$ MeV, low-energy photons cannot pass through the reactor to reach the detector by oscillating to heavy dark photons, resulting in a decrease in the signal rate of low electron recoil energy. 
Consequently, the PCGe detector with a low-energy threshold of $0.3 \sim 12.3$ keV has a weakened sensitivity in the mass range of $0.2 \sim 1$ MeV. In contrast, the CsI detector with a high-energy threshold of $3 \sim 8$ MeV, remains constant in the $0.2 \sim 1$ MeV mass range.
\\

\section{Conclusion}

In this study, we have calculated  the detection flux for the conversion probability of $\gamma \to A'$ in neutrino experiments, such as TEXONO. Unlike conventional methods that solely consider the absorption process, we also take into account the multiple scattering effect resulting from Compton scattering in the medium. Our findings indicate that this multiple Compton scattering effect significantly enhances the sensitivity to dark photons across all probed mass region. This kind of enhancement does not exhibit in the sterile neutrino production, either due to the failure to meet the scattering dominance criterion or the neutrino damping in resonant production.
In the future, the new reactor experiment, such as JUNO \cite{JUNO:2015zny} and NEOS \cite{NEOS:2015dzs} can provide better sensitivity for dark photon with the inclusion of the multiple scattering effect.
\\

\section{Acknowledgement}

We thank Zuowei Liu and Shuailiang Ge for useful discussions. 
The work of J.L. is supported by the National Science Foundation of China under Grants No. 12075005, and No. 12235001. 
The work of X.P.W. is supported by the National Science Foundation of China under Grants No. 12005009, {No. 12375095} and the Fundamental Research Funds for the Central Universities.

\section{Appendix}
\label{appendix}

\subsection{Derivation of conversion probability with multiple scattering }
\label{appsec: prob-n derivation}

In this section, we provide the derivation of Eq.~(8) in the main text, for the probability of a photon undergo $n$ times of scatterings and converting to a non-resonant dark photon.

Taking into account the damping effect of the medium (absorption and scattering), the scattering probability of a photon at differential location of $\ell\sim\ell+d\ell$ is 
\begin{equation}
d P_{\rm sca} = e^{- \Gamma \ell} (1 - e^{- \Gamma_{\rm sca} d \ell})
\simeq e^{- \Gamma \ell} \Gamma_{\rm sca} d \ell.
\end{equation} 
Thus, assuming no energy loss during medium scattering for a rough estimate, the probability of a photon being scattered once by the medium at $0< l_1 < \ell$ and then converting into a dark photon is 
\begin{equation}
P_1 (\ell) = \int_0^\ell d l_1 e^{-\Gamma l_1} \Gamma_{\rm sca} P_0 (\ell - l_1).
\end{equation}
Similarly, the conversion probability with $n$ scatterings can be obtained using the following iterative relationship
\begin{equation}
P_n (\ell) = \int_0^\ell d l_n e^{-\Gamma l_n} \Gamma_{\rm sca}
P_{n-1} (\ell - l_{n}).
\end{equation}
Here, the photon firstly scatters once with the medium at $l_n$, then scatters $n-1$ times along the remaining propagation path $\ell - l_n$ and converts into a dark photon. 
By explicitly expressing this iterative relationship, we have
\begin{widetext}
\begin{eqnarray} 
P_n (\ell)
&=& \int_0^\ell d l_n e^{-\Gamma l_n} \Gamma_{\rm sca}
\int_0^{\ell- l_n} d l_{n-1} e^{-\Gamma l_{n-1}} \Gamma_{\rm sca} 
...
\int_0^{\ell- \sum_{i=2}^n l_i} d l_1 e^{-\Gamma l_1} \Gamma_{\rm sca} 
P_0 (\ell - \sum_{i=1}^n l_i) \nonumber \\
&=& \int_0^\ell d l_n
\int_0^{\ell- l_n} d l_{n-1} 
...
\int_0^{\ell- \sum_{i=2}^n l_i} d l_1 e^{-\Gamma \sum_{i=1}^n l_i} \Gamma_{\rm sca}^n 
P_0 (\ell - \sum_{i=1}^n l_i) \nonumber \\
&=& \int_0^\ell d \Tilde{\ell}_1
\int_0^{\ell- \Tilde{\ell}_1} d \Tilde{\ell}_2 
...
\int_0^{\ell- \sum_{i=1}^{n-1} \Tilde{\ell}_i} d \Tilde{\ell}_n e^{-\Gamma \sum_{i=1}^n \Tilde{\ell}_i} \Gamma_{\rm sca}^n 
P_0 (\ell - \sum_{i=1}^n \Tilde{\ell}_i) \nonumber \\
&=& \int_{0}^\ell d \ell_1 
\int_{\ell_1}^\ell d \ell_2 ... 
\int_{\ell_{n-1}}^\ell d \ell_n
e^{-\Gamma \ell_n} \Gamma_{\rm sca}^n P_0(\ell -\ell_n). 
\label{seq: prob-n-derivation}
\end{eqnarray}
\end{widetext}
In the last two steps, we employ the following variable transformations
\begin{equation}
\Tilde{\ell}_i = l_{n+1 -i},\quad
\ell_i = \sum_{j=1}^i \Tilde{\ell}_j .
\end{equation}
The last line of Eq.~\eqref{seq: prob-n-derivation} corresponds to 
the probability 
of a photon undergoing $n$ times of scatterings
at locations $0 < \ell_1 < \ell_2, ... < \ell_n < \ell$
and converting to a non-resonant dark photon. Thus, we arrive at Eq.~(8) in the main text.

\subsection{The reactor experiment search of the dark photon}
\label{appsec: reactor search}

In this section, we briefly describe the two TEXONO detectors used in our analysis: CsI(Tl) scintillating detectors and Germanium (Ge) ionization detectors. 
We provide a detailed description on how to utilize their data to establish the dark photon limit.

The TEXONO neutrino experiment is located 28 m from the reactor core of the Kuo-Sheng Nuclear Power station with thermal power of 2.9 GW \cite{TEXONO:2000zzq}. It uses CsI(Tl) scintillating detectors \cite{TEXONO:2000zzq} and Germanium (Ge) ionization detectors \cite{TEXONO:2014eky} to search for neutrinos through $\bar{\nu}_e-e$ elastic scattering.

The CsI(Tl) scintillating crystal array measures $40 \times 31.2 \times 37~ \text{cm}^3$.
For CsI(Tl) detectors,
the energy thresholds for the electron recoil energy are $E_e^{\min} = 3$ MeV and $E_e^{\max} = 8$ MeV \cite{TEXONO:2009knm}, respectively.
We use TEXONO II-V data with a duration of 160 days.
In the typical energy range of CsI(Tl), the Compton scattering is the dominant process, 
so the probability of generating a Compton scattering signal for each dark photon is approximated by 
the conversion probability of the dark photon  \cite{Danilov:2018bks}:
\begin{equation}
P_{\text{sig}} = \epsilon^2 \frac{m_{A'}^4}{(\Delta m^2)^2}.
\label{eq:Aptogamma}
\end{equation}
Thus, the number of signal events is obtained via 
\begin{align}
N_{\text{sig}} = S T  \int dE_{A'}  \frac{1}{4\pi D^2} \frac{dF_{A'}}{dE_{A'}} P_{\text{sig}} \epsilon_{acc}(E_{A'}), 
\label{eq:NsigEvents}
\end{align}
where $S$ represents the cross-sectional area of the detector facing the reactor core, $T$ is the data collection time, and $D$ is the detector-to-reactor core distance. 
The differential flux for $A'$ is $1/(4\pi D^2) dF_{A'}/dE_{A'} $,
and after multiplying with the conversion probability $P_{\rm sig}$, $A'$ converts back into $\gamma$ without an energy change ($E_{A'} = E_\gamma$) \cite{Danilov:2018bks}.
The detector acceptance factor $\epsilon_{acc}$ reads
\begin{align}
    \epsilon_{acc}(E_{\gamma}) = \frac{1}{\sigma_C(E_{\gamma})}\int^{E_e^{\max}}_{E_e^{\min}} d E_e \frac{d\sigma_C(E_\gamma)}{dE_e},
\end{align}
where $E_\gamma$ is the initial photon energy in Compton scattering. 
The $95\%$ C.L. upper limit on the signal events from the dark photon is $1174.1$ from TEXONO data \cite{TEXONO:2009knm}.
Therefore, we require $ N_{\rm sig} < 1174.1$ to set limits in our scenario. 

TEXONO also includes an n-type PCGe detector with a of mass is 500 g
and a low electron recoil energy threshold of $0.3\sim 12.3$ keV. 
The expected events rate $N_{\text{sig}}/T$ at PCGe detector is derived from Eq.~\eqref{eq:NsigEvents} with the cross-sectional area $S$ replaced by $N_e \sigma_C(E_\gamma)$, where $N_e$ is the electron number in the detector.
This is because the size of PCGe detector is comparable to the Compton scattering interaction length (a few cm), and then we can no longer assume that every converted photon is scattered in the PCGe detector. 
We use the reactor ON/OFF data with exposure of $124.2/70.3$ kg-days \cite{TEXONO:2014eky, TEXONO:2018nir}. 
Our expected event rates take the same bins as the TEXONO PCGe data: 
60 bins with bin width 0.2 keV 
for each in the electron recoil energy from 0.3 to 12.3 keV \cite{TEXONO:2018nir}. 
At last, we use the minimal-$\chi^2$ fit method to determine the $95\%$ C.L. upper limit in our scenario.

\bibliography{ref}

\begin{thebibliography}{57}%
\makeatletter
\providecommand \@ifxundefined [1]{%
 \@ifx{#1\undefined}
}%
\providecommand \@ifnum [1]{%
 \ifnum #1\expandafter \@firstoftwo
 \else \expandafter \@secondoftwo
 \fi
}%
\providecommand \@ifx [1]{%
 \ifx #1\expandafter \@firstoftwo
 \else \expandafter \@secondoftwo
 \fi
}%
\providecommand \natexlab [1]{#1}%
\providecommand \enquote  [1]{``#1''}%
\providecommand \bibnamefont  [1]{#1}%
\providecommand \bibfnamefont [1]{#1}%
\providecommand \citenamefont [1]{#1}%
\providecommand \href@noop [0]{\@secondoftwo}%
\providecommand \href [0]{\begingroup \@sanitize@url \@href}%
\providecommand \@href[1]{\@@startlink{#1}\@@href}%
\providecommand \@@href[1]{\endgroup#1\@@endlink}%
\providecommand \@sanitize@url [0]{\catcode `\\12\catcode `\$12\catcode
  `\&12\catcode `\#12\catcode `\^12\catcode `\_12\catcode `\%12\relax}%
\providecommand \@@startlink[1]{}%
\providecommand \@@endlink[0]{}%
\providecommand \url  [0]{\begingroup\@sanitize@url \@url }%
\providecommand \@url [1]{\endgroup\@href {#1}{\urlprefix }}%
\providecommand \urlprefix  [0]{URL }%
\providecommand \Eprint [0]{\href }%
\providecommand \doibase [0]{https://doi.org/}%
\providecommand \selectlanguage [0]{\@gobble}%
\providecommand \bibinfo  [0]{\@secondoftwo}%
\providecommand \bibfield  [0]{\@secondoftwo}%
\providecommand \translation [1]{[#1]}%
\providecommand \BibitemOpen [0]{}%
\providecommand \bibitemStop [0]{}%
\providecommand \bibitemNoStop [0]{.\EOS\space}%
\providecommand \EOS [0]{\spacefactor3000\relax}%
\providecommand \BibitemShut  [1]{\csname bibitem#1\endcsname}%
\let\auto@bib@innerbib\@empty
\bibitem [{\citenamefont {Fayet}(1980)}]{FAYET1980285}%
  \BibitemOpen
  \bibfield  {author} {\bibinfo {author} {\bibfnamefont {P.}~\bibnamefont
  {Fayet}},\ }\bibfield  {title} {\bibinfo {title} {Effects of the spin-1
  partner of the goldstino (gravitino) on neutral current phenomenology},\
  }\href {https://doi.org/https://doi.org/10.1016/0370-2693(80)90488-8}
  {\bibfield  {journal} {\bibinfo  {journal} {Physics Letters B}\ }\textbf
  {\bibinfo {volume} {95}},\ \bibinfo {pages} {285} (\bibinfo {year}
  {1980})}\BibitemShut {NoStop}%
\bibitem [{\citenamefont {Fayet}(1981)}]{Fayet:1980rr}%
  \BibitemOpen
  \bibfield  {author} {\bibinfo {author} {\bibfnamefont {P.}~\bibnamefont
  {Fayet}},\ }\bibfield  {title} {\bibinfo {title} {{On the Search for a New
  Spin 1 Boson}},\ }\href {https://doi.org/10.1016/0550-3213(81)90122-X}
  {\bibfield  {journal} {\bibinfo  {journal} {Nucl. Phys. B}\ }\textbf
  {\bibinfo {volume} {187}},\ \bibinfo {pages} {184} (\bibinfo {year}
  {1981})}\BibitemShut {NoStop}%
\bibitem [{\citenamefont {Okun}(1982)}]{Okun:1982xi}%
  \BibitemOpen
  \bibfield  {author} {\bibinfo {author} {\bibfnamefont {L.~B.}\ \bibnamefont
  {Okun}},\ }\bibfield  {title} {\bibinfo {title} {{LIMITS OF ELECTRODYNAMICS:
  PARAPHOTONS?}},\ }\href@noop {} {\bibfield  {journal} {\bibinfo  {journal}
  {Sov. Phys. JETP}\ }\textbf {\bibinfo {volume} {56}},\ \bibinfo {pages} {502}
  (\bibinfo {year} {1982})}\BibitemShut {NoStop}%
\bibitem [{\citenamefont {Galison}\ and\ \citenamefont
  {Manohar}(1984)}]{Galison:1983pa}%
  \BibitemOpen
  \bibfield  {author} {\bibinfo {author} {\bibfnamefont {P.}~\bibnamefont
  {Galison}}\ and\ \bibinfo {author} {\bibfnamefont {A.}~\bibnamefont
  {Manohar}},\ }\bibfield  {title} {\bibinfo {title} {{TWO Z's OR NOT TWO
  Z's?}},\ }\href {https://doi.org/10.1016/0370-2693(84)91161-4} {\bibfield
  {journal} {\bibinfo  {journal} {Phys. Lett. B}\ }\textbf {\bibinfo {volume}
  {136}},\ \bibinfo {pages} {279} (\bibinfo {year} {1984})}\BibitemShut
  {NoStop}%
\bibitem [{\citenamefont {Holdom}(1986)}]{Holdom:1985ag}%
  \BibitemOpen
  \bibfield  {author} {\bibinfo {author} {\bibfnamefont {B.}~\bibnamefont
  {Holdom}},\ }\bibfield  {title} {\bibinfo {title} {{Two U(1)'s and Epsilon
  Charge Shifts}},\ }\href {https://doi.org/10.1016/0370-2693(86)91377-8}
  {\bibfield  {journal} {\bibinfo  {journal} {Phys. Lett. B}\ }\textbf
  {\bibinfo {volume} {166}},\ \bibinfo {pages} {196} (\bibinfo {year}
  {1986})}\BibitemShut {NoStop}%
\bibitem [{\citenamefont {Fayet}(1990)}]{Fayet:1990wx}%
  \BibitemOpen
  \bibfield  {author} {\bibinfo {author} {\bibfnamefont {P.}~\bibnamefont
  {Fayet}},\ }\bibfield  {title} {\bibinfo {title} {{Extra U(1)'s and New
  Forces}},\ }\href {https://doi.org/10.1016/0550-3213(90)90381-M} {\bibfield
  {journal} {\bibinfo  {journal} {Nucl. Phys. B}\ }\textbf {\bibinfo {volume}
  {347}},\ \bibinfo {pages} {743} (\bibinfo {year} {1990})}\BibitemShut
  {NoStop}%
\bibitem [{\citenamefont {Foot}\ and\ \citenamefont {He}(1991)}]{Foot:1991kb}%
  \BibitemOpen
  \bibfield  {author} {\bibinfo {author} {\bibfnamefont {R.}~\bibnamefont
  {Foot}}\ and\ \bibinfo {author} {\bibfnamefont {X.-G.}\ \bibnamefont {He}},\
  }\bibfield  {title} {\bibinfo {title} {{Comment on Z Z-prime mixing in
  extended gauge theories}},\ }\href
  {https://doi.org/10.1016/0370-2693(91)90901-2} {\bibfield  {journal}
  {\bibinfo  {journal} {Phys. Lett. B}\ }\textbf {\bibinfo {volume} {267}},\
  \bibinfo {pages} {509} (\bibinfo {year} {1991})}\BibitemShut {NoStop}%
\bibitem [{\citenamefont {Pospelov}\ \emph {et~al.}(2008)\citenamefont
  {Pospelov}, \citenamefont {Ritz},\ and\ \citenamefont
  {Voloshin}}]{Pospelov:2007mp}%
  \BibitemOpen
  \bibfield  {author} {\bibinfo {author} {\bibfnamefont {M.}~\bibnamefont
  {Pospelov}}, \bibinfo {author} {\bibfnamefont {A.}~\bibnamefont {Ritz}},\
  and\ \bibinfo {author} {\bibfnamefont {M.~B.}\ \bibnamefont {Voloshin}},\
  }\bibfield  {title} {\bibinfo {title} {{Secluded WIMP Dark Matter}},\ }\href
  {https://doi.org/10.1016/j.physletb.2008.02.052} {\bibfield  {journal}
  {\bibinfo  {journal} {Phys. Lett. B}\ }\textbf {\bibinfo {volume} {662}},\
  \bibinfo {pages} {53} (\bibinfo {year} {2008})},\ \Eprint
  {https://arxiv.org/abs/0711.4866} {arXiv:0711.4866 [hep-ph]} \BibitemShut
  {NoStop}%
\bibitem [{\citenamefont {Arkani-Hamed}\ \emph {et~al.}(2009)\citenamefont
  {Arkani-Hamed}, \citenamefont {Finkbeiner}, \citenamefont {Slatyer},\ and\
  \citenamefont {Weiner}}]{Arkani-Hamed:2008hhe}%
  \BibitemOpen
  \bibfield  {author} {\bibinfo {author} {\bibfnamefont {N.}~\bibnamefont
  {Arkani-Hamed}}, \bibinfo {author} {\bibfnamefont {D.~P.}\ \bibnamefont
  {Finkbeiner}}, \bibinfo {author} {\bibfnamefont {T.~R.}\ \bibnamefont
  {Slatyer}},\ and\ \bibinfo {author} {\bibfnamefont {N.}~\bibnamefont
  {Weiner}},\ }\bibfield  {title} {\bibinfo {title} {{A Theory of Dark
  Matter}},\ }\href {https://doi.org/10.1103/PhysRevD.79.015014} {\bibfield
  {journal} {\bibinfo  {journal} {Phys. Rev. D}\ }\textbf {\bibinfo {volume}
  {79}},\ \bibinfo {pages} {015014} (\bibinfo {year} {2009})},\ \Eprint
  {https://arxiv.org/abs/0810.0713} {arXiv:0810.0713 [hep-ph]} \BibitemShut
  {NoStop}%
\bibitem [{\citenamefont {Essig}\ \emph {et~al.}(2013)\citenamefont {Essig}
  \emph {et~al.}}]{Essig:2013lka}%
  \BibitemOpen
  \bibfield  {author} {\bibinfo {author} {\bibfnamefont {R.}~\bibnamefont
  {Essig}} \emph {et~al.},\ }\bibfield  {title} {\bibinfo {title} {{Working
  Group Report: New Light Weakly Coupled Particles}},\ }in\ \href@noop {}
  {\emph {\bibinfo {booktitle} {Community Summer Study 2013: Snowmass on the
  Mississippi}}}\ (\bibinfo {year} {2013})\ \Eprint
  {https://arxiv.org/abs/1311.0029} {arXiv:1311.0029 [hep-ph]} \BibitemShut
  {NoStop}%
\bibitem [{\citenamefont {Alexander}\ \emph {et~al.}(2016)\citenamefont
  {Alexander} \emph {et~al.}}]{Alexander:2016aln}%
  \BibitemOpen
  \bibfield  {author} {\bibinfo {author} {\bibfnamefont {J.}~\bibnamefont
  {Alexander}} \emph {et~al.},\ }\bibfield  {title} {\bibinfo {title} {{Dark
  Sectors 2016 Workshop: Community Report}}\ }(\bibinfo {year} {2016})\ \Eprint
  {https://arxiv.org/abs/1608.08632} {arXiv:1608.08632 [hep-ph]} \BibitemShut
  {NoStop}%
\bibitem [{\citenamefont {Battaglieri}\ \emph {et~al.}(2017)\citenamefont
  {Battaglieri} \emph {et~al.}}]{Battaglieri:2017aum}%
  \BibitemOpen
  \bibfield  {author} {\bibinfo {author} {\bibfnamefont {M.}~\bibnamefont
  {Battaglieri}} \emph {et~al.},\ }\bibfield  {title} {\bibinfo {title} {{US
  Cosmic Visions: New Ideas in Dark Matter 2017: Community Report}},\ }in\
  \href@noop {} {\emph {\bibinfo {booktitle} {{U.S. Cosmic Visions: New Ideas
  in Dark Matter}}}}\ (\bibinfo {year} {2017})\ \Eprint
  {https://arxiv.org/abs/1707.04591} {arXiv:1707.04591 [hep-ph]} \BibitemShut
  {NoStop}%
\bibitem [{\citenamefont {Redondo}\ and\ \citenamefont
  {Postma}(2009)}]{Redondo:2008ec}%
  \BibitemOpen
  \bibfield  {author} {\bibinfo {author} {\bibfnamefont {J.}~\bibnamefont
  {Redondo}}\ and\ \bibinfo {author} {\bibfnamefont {M.}~\bibnamefont
  {Postma}},\ }\bibfield  {title} {\bibinfo {title} {{Massive hidden photons as
  lukewarm dark matter}},\ }\href
  {https://doi.org/10.1088/1475-7516/2009/02/005} {\bibfield  {journal}
  {\bibinfo  {journal} {JCAP}\ }\textbf {\bibinfo {volume} {02}},\ \bibinfo
  {pages} {005}},\ \Eprint {https://arxiv.org/abs/0811.0326} {arXiv:0811.0326
  [hep-ph]} \BibitemShut {NoStop}%
\bibitem [{\citenamefont {Nelson}\ and\ \citenamefont
  {Scholtz}(2011)}]{Nelson:2011sf}%
  \BibitemOpen
  \bibfield  {author} {\bibinfo {author} {\bibfnamefont {A.~E.}\ \bibnamefont
  {Nelson}}\ and\ \bibinfo {author} {\bibfnamefont {J.}~\bibnamefont
  {Scholtz}},\ }\bibfield  {title} {\bibinfo {title} {{Dark Light, Dark Matter
  and the Misalignment Mechanism}},\ }\href
  {https://doi.org/10.1103/PhysRevD.84.103501} {\bibfield  {journal} {\bibinfo
  {journal} {Phys. Rev. D}\ }\textbf {\bibinfo {volume} {84}},\ \bibinfo
  {pages} {103501} (\bibinfo {year} {2011})},\ \Eprint
  {https://arxiv.org/abs/1105.2812} {arXiv:1105.2812 [hep-ph]} \BibitemShut
  {NoStop}%
\bibitem [{\citenamefont {Arias}\ \emph {et~al.}(2012)\citenamefont {Arias},
  \citenamefont {Cadamuro}, \citenamefont {Goodsell}, \citenamefont {Jaeckel},
  \citenamefont {Redondo},\ and\ \citenamefont {Ringwald}}]{Arias:2012az}%
  \BibitemOpen
  \bibfield  {author} {\bibinfo {author} {\bibfnamefont {P.}~\bibnamefont
  {Arias}}, \bibinfo {author} {\bibfnamefont {D.}~\bibnamefont {Cadamuro}},
  \bibinfo {author} {\bibfnamefont {M.}~\bibnamefont {Goodsell}}, \bibinfo
  {author} {\bibfnamefont {J.}~\bibnamefont {Jaeckel}}, \bibinfo {author}
  {\bibfnamefont {J.}~\bibnamefont {Redondo}},\ and\ \bibinfo {author}
  {\bibfnamefont {A.}~\bibnamefont {Ringwald}},\ }\bibfield  {title} {\bibinfo
  {title} {{WISPy Cold Dark Matter}},\ }\href
  {https://doi.org/10.1088/1475-7516/2012/06/013} {\bibfield  {journal}
  {\bibinfo  {journal} {JCAP}\ }\textbf {\bibinfo {volume} {06}},\ \bibinfo
  {pages} {013}},\ \Eprint {https://arxiv.org/abs/1201.5902} {arXiv:1201.5902
  [hep-ph]} \BibitemShut {NoStop}%
\bibitem [{\citenamefont {Graham}\ \emph {et~al.}(2016)\citenamefont {Graham},
  \citenamefont {Mardon},\ and\ \citenamefont {Rajendran}}]{Graham:2015rva}%
  \BibitemOpen
  \bibfield  {author} {\bibinfo {author} {\bibfnamefont {P.~W.}\ \bibnamefont
  {Graham}}, \bibinfo {author} {\bibfnamefont {J.}~\bibnamefont {Mardon}},\
  and\ \bibinfo {author} {\bibfnamefont {S.}~\bibnamefont {Rajendran}},\
  }\bibfield  {title} {\bibinfo {title} {{Vector Dark Matter from Inflationary
  Fluctuations}},\ }\href {https://doi.org/10.1103/PhysRevD.93.103520}
  {\bibfield  {journal} {\bibinfo  {journal} {Phys. Rev. D}\ }\textbf {\bibinfo
  {volume} {93}},\ \bibinfo {pages} {103520} (\bibinfo {year} {2016})},\
  \Eprint {https://arxiv.org/abs/1504.02102} {arXiv:1504.02102 [hep-ph]}
  \BibitemShut {NoStop}%
\bibitem [{\citenamefont {Kors}\ and\ \citenamefont
  {Nath}(2005)}]{Kors:2005uz}%
  \BibitemOpen
  \bibfield  {author} {\bibinfo {author} {\bibfnamefont {B.}~\bibnamefont
  {Kors}}\ and\ \bibinfo {author} {\bibfnamefont {P.}~\bibnamefont {Nath}},\
  }\bibfield  {title} {\bibinfo {title} {{Aspects of the Stueckelberg
  extension}},\ }\href {https://doi.org/10.1088/1126-6708/2005/07/069}
  {\bibfield  {journal} {\bibinfo  {journal} {JHEP}\ }\textbf {\bibinfo
  {volume} {07}},\ \bibinfo {pages} {069}},\ \Eprint
  {https://arxiv.org/abs/hep-ph/0503208} {arXiv:hep-ph/0503208} \BibitemShut
  {NoStop}%
\bibitem [{\citenamefont {Feldman}\ \emph
  {et~al.}(2006{\natexlab{a}})\citenamefont {Feldman}, \citenamefont {Liu},\
  and\ \citenamefont {Nath}}]{Feldman:2006ce}%
  \BibitemOpen
  \bibfield  {author} {\bibinfo {author} {\bibfnamefont {D.}~\bibnamefont
  {Feldman}}, \bibinfo {author} {\bibfnamefont {Z.}~\bibnamefont {Liu}},\ and\
  \bibinfo {author} {\bibfnamefont {P.}~\bibnamefont {Nath}},\ }\bibfield
  {title} {\bibinfo {title} {{Probing a very narrow Z-prime boson with CDF and
  D0 data}},\ }\href {https://doi.org/10.1103/PhysRevLett.97.021801} {\bibfield
   {journal} {\bibinfo  {journal} {Phys. Rev. Lett.}\ }\textbf {\bibinfo
  {volume} {97}},\ \bibinfo {pages} {021801} (\bibinfo {year}
  {2006}{\natexlab{a}})},\ \Eprint {https://arxiv.org/abs/hep-ph/0603039}
  {arXiv:hep-ph/0603039} \BibitemShut {NoStop}%
\bibitem [{\citenamefont {Feldman}\ \emph
  {et~al.}(2006{\natexlab{b}})\citenamefont {Feldman}, \citenamefont {Liu},\
  and\ \citenamefont {Nath}}]{Feldman:2006wb}%
  \BibitemOpen
  \bibfield  {author} {\bibinfo {author} {\bibfnamefont {D.}~\bibnamefont
  {Feldman}}, \bibinfo {author} {\bibfnamefont {Z.}~\bibnamefont {Liu}},\ and\
  \bibinfo {author} {\bibfnamefont {P.}~\bibnamefont {Nath}},\ }\bibfield
  {title} {\bibinfo {title} {{The Stueckelberg $Z$ Prime at the LHC: Discovery
  Potential, Signature Spaces and Model Discrimination}},\ }\href
  {https://doi.org/10.1088/1126-6708/2006/11/007} {\bibfield  {journal}
  {\bibinfo  {journal} {JHEP}\ }\textbf {\bibinfo {volume} {11}},\ \bibinfo
  {pages} {007}},\ \Eprint {https://arxiv.org/abs/hep-ph/0606294}
  {arXiv:hep-ph/0606294} \BibitemShut {NoStop}%
\bibitem [{\citenamefont {Feldman}\ \emph {et~al.}(2007)\citenamefont
  {Feldman}, \citenamefont {Liu},\ and\ \citenamefont {Nath}}]{Feldman:2007wj}%
  \BibitemOpen
  \bibfield  {author} {\bibinfo {author} {\bibfnamefont {D.}~\bibnamefont
  {Feldman}}, \bibinfo {author} {\bibfnamefont {Z.}~\bibnamefont {Liu}},\ and\
  \bibinfo {author} {\bibfnamefont {P.}~\bibnamefont {Nath}},\ }\bibfield
  {title} {\bibinfo {title} {{The Stueckelberg Z-prime Extension with Kinetic
  Mixing and Milli-Charged Dark Matter From the Hidden Sector}},\ }\href
  {https://doi.org/10.1103/PhysRevD.75.115001} {\bibfield  {journal} {\bibinfo
  {journal} {Phys. Rev. D}\ }\textbf {\bibinfo {volume} {75}},\ \bibinfo
  {pages} {115001} (\bibinfo {year} {2007})},\ \Eprint
  {https://arxiv.org/abs/hep-ph/0702123} {arXiv:hep-ph/0702123} \BibitemShut
  {NoStop}%
\bibitem [{\citenamefont {Feldman}\ \emph {et~al.}(2009)\citenamefont
  {Feldman}, \citenamefont {Liu}, \citenamefont {Nath},\ and\ \citenamefont
  {Nelson}}]{Feldman:2009wv}%
  \BibitemOpen
  \bibfield  {author} {\bibinfo {author} {\bibfnamefont {D.}~\bibnamefont
  {Feldman}}, \bibinfo {author} {\bibfnamefont {Z.}~\bibnamefont {Liu}},
  \bibinfo {author} {\bibfnamefont {P.}~\bibnamefont {Nath}},\ and\ \bibinfo
  {author} {\bibfnamefont {B.~D.}\ \bibnamefont {Nelson}},\ }\bibfield  {title}
  {\bibinfo {title} {{Explaining PAMELA and WMAP data through Coannihilations
  in Extended SUGRA with Collider Implications}},\ }\href
  {https://doi.org/10.1103/PhysRevD.80.075001} {\bibfield  {journal} {\bibinfo
  {journal} {Phys. Rev. D}\ }\textbf {\bibinfo {volume} {80}},\ \bibinfo
  {pages} {075001} (\bibinfo {year} {2009})},\ \Eprint
  {https://arxiv.org/abs/0907.5392} {arXiv:0907.5392 [hep-ph]} \BibitemShut
  {NoStop}%
\bibitem [{\citenamefont {Du}\ \emph {et~al.}(2020)\citenamefont {Du},
  \citenamefont {Liu},\ and\ \citenamefont {Tran}}]{Du:2019mlc}%
  \BibitemOpen
  \bibfield  {author} {\bibinfo {author} {\bibfnamefont {M.}~\bibnamefont
  {Du}}, \bibinfo {author} {\bibfnamefont {Z.}~\bibnamefont {Liu}},\ and\
  \bibinfo {author} {\bibfnamefont {V.~Q.}\ \bibnamefont {Tran}},\ }\bibfield
  {title} {\bibinfo {title} {{Enhanced Long-Lived Dark Photon Signals at the
  LHC}},\ }\href {https://doi.org/10.1007/JHEP05(2020)055} {\bibfield
  {journal} {\bibinfo  {journal} {JHEP}\ }\textbf {\bibinfo {volume} {05}},\
  \bibinfo {pages} {055}},\ \Eprint {https://arxiv.org/abs/1912.00422}
  {arXiv:1912.00422 [hep-ph]} \BibitemShut {NoStop}%
\bibitem [{\citenamefont {Du}\ \emph {et~al.}(2022)\citenamefont {Du},
  \citenamefont {Fang}, \citenamefont {Liu},\ and\ \citenamefont
  {Tran}}]{Du:2021cmt}%
  \BibitemOpen
  \bibfield  {author} {\bibinfo {author} {\bibfnamefont {M.}~\bibnamefont
  {Du}}, \bibinfo {author} {\bibfnamefont {R.}~\bibnamefont {Fang}}, \bibinfo
  {author} {\bibfnamefont {Z.}~\bibnamefont {Liu}},\ and\ \bibinfo {author}
  {\bibfnamefont {V.~Q.}\ \bibnamefont {Tran}},\ }\bibfield  {title} {\bibinfo
  {title} {{Enhanced long-lived dark photon signals at lifetime frontier
  detectors}},\ }\href {https://doi.org/10.1103/PhysRevD.105.055012} {\bibfield
   {journal} {\bibinfo  {journal} {Phys. Rev. D}\ }\textbf {\bibinfo {volume}
  {105}},\ \bibinfo {pages} {055012} (\bibinfo {year} {2022})},\ \Eprint
  {https://arxiv.org/abs/2111.15503} {arXiv:2111.15503 [hep-ph]} \BibitemShut
  {NoStop}%
\bibitem [{\citenamefont {Redi}\ and\ \citenamefont
  {Tesi}(2022)}]{Redi:2022zkt}%
  \BibitemOpen
  \bibfield  {author} {\bibinfo {author} {\bibfnamefont {M.}~\bibnamefont
  {Redi}}\ and\ \bibinfo {author} {\bibfnamefont {A.}~\bibnamefont {Tesi}},\
  }\bibfield  {title} {\bibinfo {title} {{Dark photon Dark Matter without
  Stueckelberg mass}},\ }\href {https://doi.org/10.1007/JHEP10(2022)167}
  {\bibfield  {journal} {\bibinfo  {journal} {JHEP}\ }\textbf {\bibinfo
  {volume} {10}},\ \bibinfo {pages} {167}},\ \Eprint
  {https://arxiv.org/abs/2204.14274} {arXiv:2204.14274 [hep-ph]} \BibitemShut
  {NoStop}%
\bibitem [{\citenamefont {Fabbrichesi}\ \emph {et~al.}(2020)\citenamefont
  {Fabbrichesi}, \citenamefont {Gabrielli},\ and\ \citenamefont
  {Lanfranchi}}]{Fabbrichesi:2020wbt}%
  \BibitemOpen
  \bibfield  {author} {\bibinfo {author} {\bibfnamefont {M.}~\bibnamefont
  {Fabbrichesi}}, \bibinfo {author} {\bibfnamefont {E.}~\bibnamefont
  {Gabrielli}},\ and\ \bibinfo {author} {\bibfnamefont {G.}~\bibnamefont
  {Lanfranchi}},\ }\bibfield  {title} {\bibinfo {title} {{The Dark Photon}}\
  }\href {https://doi.org/10.1007/978-3-030-62519-1}
  {10.1007/978-3-030-62519-1} (\bibinfo {year} {2020}),\ \Eprint
  {https://arxiv.org/abs/2005.01515} {arXiv:2005.01515 [hep-ph]} \BibitemShut
  {NoStop}%
\bibitem [{\citenamefont {Caputo}\ \emph {et~al.}(2021)\citenamefont {Caputo},
  \citenamefont {Millar}, \citenamefont {O'Hare},\ and\ \citenamefont
  {Vitagliano}}]{Caputo:2021eaa}%
  \BibitemOpen
  \bibfield  {author} {\bibinfo {author} {\bibfnamefont {A.}~\bibnamefont
  {Caputo}}, \bibinfo {author} {\bibfnamefont {A.~J.}\ \bibnamefont {Millar}},
  \bibinfo {author} {\bibfnamefont {C.~A.~J.}\ \bibnamefont {O'Hare}},\ and\
  \bibinfo {author} {\bibfnamefont {E.}~\bibnamefont {Vitagliano}},\ }\bibfield
   {title} {\bibinfo {title} {{Dark photon limits: A handbook}},\ }\href
  {https://doi.org/10.1103/PhysRevD.104.095029} {\bibfield  {journal} {\bibinfo
   {journal} {Phys. Rev. D}\ }\textbf {\bibinfo {volume} {104}},\ \bibinfo
  {pages} {095029} (\bibinfo {year} {2021})},\ \Eprint
  {https://arxiv.org/abs/2105.04565} {arXiv:2105.04565 [hep-ph]} \BibitemShut
  {NoStop}%
\bibitem [{\citenamefont {Demidov}\ \emph {et~al.}(2019)\citenamefont
  {Demidov}, \citenamefont {Gninenko},\ and\ \citenamefont
  {Gorbunov}}]{Demidov:2018odn}%
  \BibitemOpen
  \bibfield  {author} {\bibinfo {author} {\bibfnamefont {S.}~\bibnamefont
  {Demidov}}, \bibinfo {author} {\bibfnamefont {S.}~\bibnamefont {Gninenko}},\
  and\ \bibinfo {author} {\bibfnamefont {D.}~\bibnamefont {Gorbunov}},\
  }\bibfield  {title} {\bibinfo {title} {{Light hidden photon production in
  high energy collisions}},\ }\href {https://doi.org/10.1007/JHEP07(2019)162}
  {\bibfield  {journal} {\bibinfo  {journal} {JHEP}\ }\textbf {\bibinfo
  {volume} {07}},\ \bibinfo {pages} {162}},\ \Eprint
  {https://arxiv.org/abs/1812.02719} {arXiv:1812.02719 [hep-ph]} \BibitemShut
  {NoStop}%
\bibitem [{\citenamefont {Ilten}\ \emph {et~al.}(2018)\citenamefont {Ilten},
  \citenamefont {Soreq}, \citenamefont {Williams},\ and\ \citenamefont
  {Xue}}]{Ilten:2018crw}%
  \BibitemOpen
  \bibfield  {author} {\bibinfo {author} {\bibfnamefont {P.}~\bibnamefont
  {Ilten}}, \bibinfo {author} {\bibfnamefont {Y.}~\bibnamefont {Soreq}},
  \bibinfo {author} {\bibfnamefont {M.}~\bibnamefont {Williams}},\ and\
  \bibinfo {author} {\bibfnamefont {W.}~\bibnamefont {Xue}},\ }\bibfield
  {title} {\bibinfo {title} {{Serendipity in dark photon searches}},\ }\href
  {https://doi.org/10.1007/JHEP06(2018)004} {\bibfield  {journal} {\bibinfo
  {journal} {JHEP}\ }\textbf {\bibinfo {volume} {06}},\ \bibinfo {pages}
  {004}},\ \Eprint {https://arxiv.org/abs/1801.04847} {arXiv:1801.04847
  [hep-ph]} \BibitemShut {NoStop}%
\bibitem [{\citenamefont {Bauer}\ \emph {et~al.}(2018)\citenamefont {Bauer},
  \citenamefont {Foldenauer},\ and\ \citenamefont {Jaeckel}}]{Bauer:2018onh}%
  \BibitemOpen
  \bibfield  {author} {\bibinfo {author} {\bibfnamefont {M.}~\bibnamefont
  {Bauer}}, \bibinfo {author} {\bibfnamefont {P.}~\bibnamefont {Foldenauer}},\
  and\ \bibinfo {author} {\bibfnamefont {J.}~\bibnamefont {Jaeckel}},\
  }\bibfield  {title} {\bibinfo {title} {{Hunting All the Hidden Photons}},\
  }\href {https://doi.org/10.1007/JHEP07(2018)094} {\bibfield  {journal}
  {\bibinfo  {journal} {JHEP}\ }\textbf {\bibinfo {volume} {07}},\ \bibinfo
  {pages} {094}},\ \Eprint {https://arxiv.org/abs/1803.05466} {arXiv:1803.05466
  [hep-ph]} \BibitemShut {NoStop}%
\bibitem [{\citenamefont {Danilov}\ \emph {et~al.}(2019)\citenamefont
  {Danilov}, \citenamefont {Demidov},\ and\ \citenamefont
  {Gorbunov}}]{Danilov:2018bks}%
  \BibitemOpen
  \bibfield  {author} {\bibinfo {author} {\bibfnamefont {M.}~\bibnamefont
  {Danilov}}, \bibinfo {author} {\bibfnamefont {S.}~\bibnamefont {Demidov}},\
  and\ \bibinfo {author} {\bibfnamefont {D.}~\bibnamefont {Gorbunov}},\
  }\bibfield  {title} {\bibinfo {title} {{Constraints on hidden photons
  produced in nuclear reactors}},\ }\href
  {https://doi.org/10.1103/PhysRevLett.122.041801} {\bibfield  {journal}
  {\bibinfo  {journal} {Phys. Rev. Lett.}\ }\textbf {\bibinfo {volume} {122}},\
  \bibinfo {pages} {041801} (\bibinfo {year} {2019})},\ \Eprint
  {https://arxiv.org/abs/1804.10777} {arXiv:1804.10777 [hep-ph]} \BibitemShut
  {NoStop}%
\bibitem [{\citenamefont {Smirnov}\ \emph {et~al.}(2021)\citenamefont
  {Smirnov}, \citenamefont {Yang}, \citenamefont {Liao}, \citenamefont {Hu},\
  and\ \citenamefont {Ling}}]{Smirnov:2021wgi}%
  \BibitemOpen
  \bibfield  {author} {\bibinfo {author} {\bibfnamefont {M.}~\bibnamefont
  {Smirnov}}, \bibinfo {author} {\bibfnamefont {G.}~\bibnamefont {Yang}},
  \bibinfo {author} {\bibfnamefont {J.}~\bibnamefont {Liao}}, \bibinfo {author}
  {\bibfnamefont {Z.}~\bibnamefont {Hu}},\ and\ \bibinfo {author}
  {\bibfnamefont {J.}~\bibnamefont {Ling}},\ }\bibfield  {title} {\bibinfo
  {title} {{Light dark bosons in the JUNO-TAO neutrino detector}},\ }\href
  {https://doi.org/10.1103/PhysRevD.104.116024} {\bibfield  {journal} {\bibinfo
   {journal} {Phys. Rev. D}\ }\textbf {\bibinfo {volume} {104}},\ \bibinfo
  {pages} {116024} (\bibinfo {year} {2021})},\ \Eprint
  {https://arxiv.org/abs/2109.04276} {arXiv:2109.04276 [hep-ex]} \BibitemShut
  {NoStop}%
\bibitem [{\citenamefont {Mizumoto}\ \emph {et~al.}(2013)\citenamefont
  {Mizumoto}, \citenamefont {Ohta}, \citenamefont {Horie}, \citenamefont
  {Suzuki}, \citenamefont {Inoue},\ and\ \citenamefont
  {Minowa}}]{Mizumoto:2013jy}%
  \BibitemOpen
  \bibfield  {author} {\bibinfo {author} {\bibfnamefont {T.}~\bibnamefont
  {Mizumoto}}, \bibinfo {author} {\bibfnamefont {R.}~\bibnamefont {Ohta}},
  \bibinfo {author} {\bibfnamefont {T.}~\bibnamefont {Horie}}, \bibinfo
  {author} {\bibfnamefont {J.}~\bibnamefont {Suzuki}}, \bibinfo {author}
  {\bibfnamefont {Y.}~\bibnamefont {Inoue}},\ and\ \bibinfo {author}
  {\bibfnamefont {M.}~\bibnamefont {Minowa}},\ }\bibfield  {title} {\bibinfo
  {title} {{Experimental search for solar hidden photons in the eV energy range
  using kinetic mixing with photons}},\ }\href
  {https://doi.org/10.1088/1475-7516/2013/07/013} {\bibfield  {journal}
  {\bibinfo  {journal} {JCAP}\ }\textbf {\bibinfo {volume} {07}},\ \bibinfo
  {pages} {013}},\ \Eprint {https://arxiv.org/abs/1302.1000} {arXiv:1302.1000
  [astro-ph.SR]} \BibitemShut {NoStop}%
\bibitem [{\citenamefont {Fortin}\ and\ \citenamefont
  {Sinha}(2019)}]{Fortin:2019npr}%
  \BibitemOpen
  \bibfield  {author} {\bibinfo {author} {\bibfnamefont {J.-F.}\ \bibnamefont
  {Fortin}}\ and\ \bibinfo {author} {\bibfnamefont {K.}~\bibnamefont {Sinha}},\
  }\bibfield  {title} {\bibinfo {title} {{Photon-dark photon conversions in
  extreme background electromagnetic fields}},\ }\href
  {https://doi.org/10.1088/1475-7516/2019/11/020} {\bibfield  {journal}
  {\bibinfo  {journal} {JCAP}\ }\textbf {\bibinfo {volume} {11}},\ \bibinfo
  {pages} {020}},\ \Eprint {https://arxiv.org/abs/1904.08968} {arXiv:1904.08968
  [hep-ph]} \BibitemShut {NoStop}%
\bibitem [{\citenamefont {Workman}\ \emph {et~al.}(2022)\citenamefont {Workman}
  \emph {et~al.}}]{ParticleDataGroup:2022pth}%
  \BibitemOpen
  \bibfield  {author} {\bibinfo {author} {\bibfnamefont {R.~L.}\ \bibnamefont
  {Workman}} \emph {et~al.} (\bibinfo {collaboration} {Particle Data Group}),\
  }\bibfield  {title} {\bibinfo {title} {{Review of Particle Physics}},\ }\href
  {https://doi.org/10.1093/ptep/ptac097} {\bibfield  {journal} {\bibinfo
  {journal} {PTEP}\ }\textbf {\bibinfo {volume} {2022}},\ \bibinfo {pages}
  {083C01} (\bibinfo {year} {2022})}\BibitemShut {NoStop}%
\bibitem [{\citenamefont {Seo}\ and\ \citenamefont {Kim}(2021)}]{Seo:2020dtx}%
  \BibitemOpen
  \bibfield  {author} {\bibinfo {author} {\bibfnamefont {S.~H.}\ \bibnamefont
  {Seo}}\ and\ \bibinfo {author} {\bibfnamefont {Y.~D.}\ \bibnamefont {Kim}},\
  }\bibfield  {title} {\bibinfo {title} {{Dark Photon Search at Yemilab,
  Korea}},\ }\href {https://doi.org/10.1007/JHEP04(2021)135} {\bibfield
  {journal} {\bibinfo  {journal} {JHEP}\ }\textbf {\bibinfo {volume} {04}},\
  \bibinfo {pages} {135}},\ \Eprint {https://arxiv.org/abs/2009.11155}
  {arXiv:2009.11155 [hep-ph]} \BibitemShut {NoStop}%
\bibitem [{\citenamefont {Redondo}\ and\ \citenamefont
  {Raffelt}(2013)}]{Redondo:2013lna}%
  \BibitemOpen
  \bibfield  {author} {\bibinfo {author} {\bibfnamefont {J.}~\bibnamefont
  {Redondo}}\ and\ \bibinfo {author} {\bibfnamefont {G.}~\bibnamefont
  {Raffelt}},\ }\bibfield  {title} {\bibinfo {title} {{Solar constraints on
  hidden photons re-visited}},\ }\href
  {https://doi.org/10.1088/1475-7516/2013/08/034} {\bibfield  {journal}
  {\bibinfo  {journal} {JCAP}\ }\textbf {\bibinfo {volume} {08}},\ \bibinfo
  {pages} {034}},\ \Eprint {https://arxiv.org/abs/1305.2920} {arXiv:1305.2920
  [hep-ph]} \BibitemShut {NoStop}%
\bibitem [{\citenamefont {Redondo}(2015)}]{Redondo:2015iea}%
  \BibitemOpen
  \bibfield  {author} {\bibinfo {author} {\bibfnamefont {J.}~\bibnamefont
  {Redondo}},\ }\bibfield  {title} {\bibinfo {title} {{Atlas of solar hidden
  photon emission}},\ }\href {https://doi.org/10.1088/1475-7516/2015/07/024}
  {\bibfield  {journal} {\bibinfo  {journal} {JCAP}\ }\textbf {\bibinfo
  {volume} {07}},\ \bibinfo {pages} {024}},\ \Eprint
  {https://arxiv.org/abs/1501.07292} {arXiv:1501.07292 [hep-ph]} \BibitemShut
  {NoStop}%
\bibitem [{\citenamefont {Dent}\ \emph {et~al.}(2020)\citenamefont {Dent},
  \citenamefont {Dutta}, \citenamefont {Kim}, \citenamefont {Liao},
  \citenamefont {Mahapatra}, \citenamefont {Sinha},\ and\ \citenamefont
  {Thompson}}]{Dent:2019ueq}%
  \BibitemOpen
  \bibfield  {author} {\bibinfo {author} {\bibfnamefont {J.~B.}\ \bibnamefont
  {Dent}}, \bibinfo {author} {\bibfnamefont {B.}~\bibnamefont {Dutta}},
  \bibinfo {author} {\bibfnamefont {D.}~\bibnamefont {Kim}}, \bibinfo {author}
  {\bibfnamefont {S.}~\bibnamefont {Liao}}, \bibinfo {author} {\bibfnamefont
  {R.}~\bibnamefont {Mahapatra}}, \bibinfo {author} {\bibfnamefont
  {K.}~\bibnamefont {Sinha}},\ and\ \bibinfo {author} {\bibfnamefont
  {A.}~\bibnamefont {Thompson}},\ }\bibfield  {title} {\bibinfo {title} {{New
  Directions for Axion Searches via Scattering at Reactor Neutrino
  Experiments}},\ }\href {https://doi.org/10.1103/PhysRevLett.124.211804}
  {\bibfield  {journal} {\bibinfo  {journal} {Phys. Rev. Lett.}\ }\textbf
  {\bibinfo {volume} {124}},\ \bibinfo {pages} {211804} (\bibinfo {year}
  {2020})},\ \Eprint {https://arxiv.org/abs/1912.05733} {arXiv:1912.05733
  [hep-ph]} \BibitemShut {NoStop}%
\bibitem [{\citenamefont {M.J.~Berger}\ and\ \citenamefont
  {Olsen}()}]{Berger:xcom}%
  \BibitemOpen
  \bibfield  {author} {\bibinfo {author} {\bibfnamefont {S.~S. J. C. J. C. R.
  S. D.~Z.}\ \bibnamefont {M.J.~Berger}, \bibfnamefont {J.H.~Hubbell}}\ and\
  \bibinfo {author} {\bibfnamefont {K.}~\bibnamefont {Olsen}},\ }\href@noop {}
  {\bibinfo {title} {Xcom: Photon cross sections database}},\ \bibinfo
  {howpublished}
  {\url{https://www.nist.gov/pml/xcom-photon-cross-sections-database}}\BibitemShut
  {NoStop}%
\bibitem [{\citenamefont {Redondo}(2008)}]{Redondo:2008aa}%
  \BibitemOpen
  \bibfield  {author} {\bibinfo {author} {\bibfnamefont {J.}~\bibnamefont
  {Redondo}},\ }\bibfield  {title} {\bibinfo {title} {{Helioscope Bounds on
  Hidden Sector Photons}},\ }\href
  {https://doi.org/10.1088/1475-7516/2008/07/008} {\bibfield  {journal}
  {\bibinfo  {journal} {JCAP}\ }\textbf {\bibinfo {volume} {07}},\ \bibinfo
  {pages} {008}},\ \Eprint {https://arxiv.org/abs/0801.1527} {arXiv:0801.1527
  [hep-ph]} \BibitemShut {NoStop}%
\bibitem [{\citenamefont {An}\ \emph {et~al.}(2013)\citenamefont {An},
  \citenamefont {Pospelov},\ and\ \citenamefont {Pradler}}]{An:2013yfc}%
  \BibitemOpen
  \bibfield  {author} {\bibinfo {author} {\bibfnamefont {H.}~\bibnamefont
  {An}}, \bibinfo {author} {\bibfnamefont {M.}~\bibnamefont {Pospelov}},\ and\
  \bibinfo {author} {\bibfnamefont {J.}~\bibnamefont {Pradler}},\ }\bibfield
  {title} {\bibinfo {title} {{New stellar constraints on dark photons}},\
  }\href {https://doi.org/10.1016/j.physletb.2013.07.008} {\bibfield  {journal}
  {\bibinfo  {journal} {Phys. Lett. B}\ }\textbf {\bibinfo {volume} {725}},\
  \bibinfo {pages} {190} (\bibinfo {year} {2013})},\ \Eprint
  {https://arxiv.org/abs/1302.3884} {arXiv:1302.3884 [hep-ph]} \BibitemShut
  {NoStop}%
\bibitem [{\citenamefont {H.~Bechteler}\ and\ \citenamefont
  {Yogeshwar}(1984)}]{Bechteler:1984}%
  \BibitemOpen
  \bibfield  {author} {\bibinfo {author} {\bibfnamefont {H.~S.}\ \bibnamefont
  {H.~Bechteler}, \bibfnamefont {H.~Fasissner}}\ and\ \bibinfo {author}
  {\bibnamefont {Yogeshwar}},\ }\href@noop {} {\emph {\bibinfo {title} {{IKP
  Annual report}}}}\ (\bibinfo {year} {1984})\BibitemShut {NoStop}%
\bibitem [{\citenamefont {Stodolsky}(1987)}]{PhysRevD.36.2273}%
  \BibitemOpen
  \bibfield  {author} {\bibinfo {author} {\bibfnamefont {L.}~\bibnamefont
  {Stodolsky}},\ }\bibfield  {title} {\bibinfo {title} {Treatment of neutrino
  oscillations in a thermal environment},\ }\href
  {https://doi.org/10.1103/PhysRevD.36.2273} {\bibfield  {journal} {\bibinfo
  {journal} {Phys. Rev. D}\ }\textbf {\bibinfo {volume} {36}},\ \bibinfo
  {pages} {2273} (\bibinfo {year} {1987})}\BibitemShut {NoStop}%
\bibitem [{\citenamefont {Raffelt}\ and\ \citenamefont
  {Zhou}(2011)}]{PhysRevD.83.093014}%
  \BibitemOpen
  \bibfield  {author} {\bibinfo {author} {\bibfnamefont {G.~G.}\ \bibnamefont
  {Raffelt}}\ and\ \bibinfo {author} {\bibfnamefont {S.}~\bibnamefont {Zhou}},\
  }\bibfield  {title} {\bibinfo {title} {Supernova bound on kev-mass sterile
  neutrinos reexamined},\ }\href {https://doi.org/10.1103/PhysRevD.83.093014}
  {\bibfield  {journal} {\bibinfo  {journal} {Phys. Rev. D}\ }\textbf {\bibinfo
  {volume} {83}},\ \bibinfo {pages} {093014} (\bibinfo {year}
  {2011})}\BibitemShut {NoStop}%
\bibitem [{\citenamefont {Arg\"uelles}\ \emph {et~al.}(2019)\citenamefont
  {Arg\"uelles}, \citenamefont {Brdar},\ and\ \citenamefont
  {Kopp}}]{PhysRevD.99.043012}%
  \BibitemOpen
  \bibfield  {author} {\bibinfo {author} {\bibfnamefont {C.~A.}\ \bibnamefont
  {Arg\"uelles}}, \bibinfo {author} {\bibfnamefont {V.}~\bibnamefont {Brdar}},\
  and\ \bibinfo {author} {\bibfnamefont {J.}~\bibnamefont {Kopp}},\ }\bibfield
  {title} {\bibinfo {title} {Production of kev sterile neutrinos in supernovae:
  New constraints and gamma-ray observables},\ }\href
  {https://doi.org/10.1103/PhysRevD.99.043012} {\bibfield  {journal} {\bibinfo
  {journal} {Phys. Rev. D}\ }\textbf {\bibinfo {volume} {99}},\ \bibinfo
  {pages} {043012} (\bibinfo {year} {2019})}\BibitemShut {NoStop}%
\bibitem [{\citenamefont {Deniz}\ \emph {et~al.}(2010)\citenamefont {Deniz}
  \emph {et~al.}}]{TEXONO:2009knm}%
  \BibitemOpen
  \bibfield  {author} {\bibinfo {author} {\bibfnamefont {M.}~\bibnamefont
  {Deniz}} \emph {et~al.} (\bibinfo {collaboration} {TEXONO}),\ }\bibfield
  {title} {\bibinfo {title} {{Measurement of Nu(e)-bar -Electron Scattering
  Cross-Section with a CsI(Tl) Scintillating Crystal Array at the Kuo-Sheng
  Nuclear Power Reactor}},\ }\href {https://doi.org/10.1103/PhysRevD.81.072001}
  {\bibfield  {journal} {\bibinfo  {journal} {Phys. Rev. D}\ }\textbf {\bibinfo
  {volume} {81}},\ \bibinfo {pages} {072001} (\bibinfo {year} {2010})},\
  \Eprint {https://arxiv.org/abs/0911.1597} {arXiv:0911.1597 [hep-ex]}
  \BibitemShut {NoStop}%
\bibitem [{\citenamefont {Soma}\ \emph {et~al.}(2016)\citenamefont {Soma} \emph
  {et~al.}}]{TEXONO:2014eky}%
  \BibitemOpen
  \bibfield  {author} {\bibinfo {author} {\bibfnamefont {A.~K.}\ \bibnamefont
  {Soma}} \emph {et~al.} (\bibinfo {collaboration} {TEXONO}),\ }\bibfield
  {title} {\bibinfo {title} {{Characterization and Performance of Germanium
  Detectors with sub-keV Sensitivities for Neutrino and Dark Matter
  Experiments}},\ }\href {https://doi.org/10.1016/j.nima.2016.08.044}
  {\bibfield  {journal} {\bibinfo  {journal} {Nucl. Instrum. Meth. A}\ }\textbf
  {\bibinfo {volume} {836}},\ \bibinfo {pages} {67} (\bibinfo {year} {2016})},\
  \Eprint {https://arxiv.org/abs/1411.4802} {arXiv:1411.4802 [physics.ins-det]}
  \BibitemShut {NoStop}%
\bibitem [{\citenamefont {Singh}\ \emph {et~al.}(2019)\citenamefont {Singh}
  \emph {et~al.}}]{TEXONO:2018nir}%
  \BibitemOpen
  \bibfield  {author} {\bibinfo {author} {\bibfnamefont {L.}~\bibnamefont
  {Singh}} \emph {et~al.} (\bibinfo {collaboration} {TEXONO}),\ }\bibfield
  {title} {\bibinfo {title} {{Constraints on millicharged particles with low
  threshold germanium detectors at Kuo-Sheng Reactor Neutrino Laboratory}},\
  }\href {https://doi.org/10.1103/PhysRevD.99.032009} {\bibfield  {journal}
  {\bibinfo  {journal} {Phys. Rev. D}\ }\textbf {\bibinfo {volume} {99}},\
  \bibinfo {pages} {032009} (\bibinfo {year} {2019})},\ \Eprint
  {https://arxiv.org/abs/1808.02719} {arXiv:1808.02719 [hep-ph]} \BibitemShut
  {NoStop}%
\bibitem [{\citenamefont {Li}\ \emph {et~al.}(2001)\citenamefont {Li} \emph
  {et~al.}}]{TEXONO:2000zzq}%
  \BibitemOpen
  \bibfield  {author} {\bibinfo {author} {\bibfnamefont {H.~B.}\ \bibnamefont
  {Li}} \emph {et~al.} (\bibinfo {collaboration} {TEXONO}),\ }\bibfield
  {title} {\bibinfo {title} {{A CsI(Tl) scintillating crystal detector for the
  studies of low-energy neutrino interactions}},\ }\href
  {https://doi.org/10.1016/S0168-9002(00)00999-2} {\bibfield  {journal}
  {\bibinfo  {journal} {Nucl. Instrum. Meth. A}\ }\textbf {\bibinfo {volume}
  {459}},\ \bibinfo {pages} {93} (\bibinfo {year} {2001})},\ \bibinfo {note}
  {[Erratum: Nucl.Instrum.Meth.A 485, 821 (2002)]},\ \Eprint
  {https://arxiv.org/abs/hep-ex/0001001} {arXiv:hep-ex/0001001} \BibitemShut
  {NoStop}%
\bibitem [{\citenamefont {Jaeckel}\ and\ \citenamefont
  {Roy}(2010)}]{Jaeckel:2010xx}%
  \BibitemOpen
  \bibfield  {author} {\bibinfo {author} {\bibfnamefont {J.}~\bibnamefont
  {Jaeckel}}\ and\ \bibinfo {author} {\bibfnamefont {S.}~\bibnamefont {Roy}},\
  }\bibfield  {title} {\bibinfo {title} {{Spectroscopy as a test of Coulomb's
  law: A Probe of the hidden sector}},\ }\href
  {https://doi.org/10.1103/PhysRevD.82.125020} {\bibfield  {journal} {\bibinfo
  {journal} {Phys. Rev. D}\ }\textbf {\bibinfo {volume} {82}},\ \bibinfo
  {pages} {125020} (\bibinfo {year} {2010})},\ \Eprint
  {https://arxiv.org/abs/1008.3536} {arXiv:1008.3536 [hep-ph]} \BibitemShut
  {NoStop}%
\bibitem [{\citenamefont {Inada}\ \emph {et~al.}(2013)\citenamefont {Inada},
  \citenamefont {Namba}, \citenamefont {Asai}, \citenamefont {Kobayashi},
  \citenamefont {Tanaka}, \citenamefont {Tamasaku}, \citenamefont {Sawada},\
  and\ \citenamefont {Ishikawa}}]{INADA2013301}%
  \BibitemOpen
  \bibfield  {author} {\bibinfo {author} {\bibfnamefont {T.}~\bibnamefont
  {Inada}}, \bibinfo {author} {\bibfnamefont {T.}~\bibnamefont {Namba}},
  \bibinfo {author} {\bibfnamefont {S.}~\bibnamefont {Asai}}, \bibinfo {author}
  {\bibfnamefont {T.}~\bibnamefont {Kobayashi}}, \bibinfo {author}
  {\bibfnamefont {Y.}~\bibnamefont {Tanaka}}, \bibinfo {author} {\bibfnamefont
  {K.}~\bibnamefont {Tamasaku}}, \bibinfo {author} {\bibfnamefont
  {K.}~\bibnamefont {Sawada}},\ and\ \bibinfo {author} {\bibfnamefont
  {T.}~\bibnamefont {Ishikawa}},\ }\bibfield  {title} {\bibinfo {title}
  {Results of a search for paraphotons with intense x-ray beams at spring-8},\
  }\href {https://doi.org/https://doi.org/10.1016/j.physletb.2013.04.033}
  {\bibfield  {journal} {\bibinfo  {journal} {Physics Letters B}\ }\textbf
  {\bibinfo {volume} {722}},\ \bibinfo {pages} {301} (\bibinfo {year}
  {2013})}\BibitemShut {NoStop}%
\bibitem [{\citenamefont {Vinyoles}\ \emph {et~al.}(2015)\citenamefont
  {Vinyoles}, \citenamefont {Serenelli}, \citenamefont {Villante},
  \citenamefont {Basu}, \citenamefont {Redondo},\ and\ \citenamefont
  {Isern}}]{Vinyoles:2015aba}%
  \BibitemOpen
  \bibfield  {author} {\bibinfo {author} {\bibfnamefont {N.}~\bibnamefont
  {Vinyoles}}, \bibinfo {author} {\bibfnamefont {A.}~\bibnamefont {Serenelli}},
  \bibinfo {author} {\bibfnamefont {F.~L.}\ \bibnamefont {Villante}}, \bibinfo
  {author} {\bibfnamefont {S.}~\bibnamefont {Basu}}, \bibinfo {author}
  {\bibfnamefont {J.}~\bibnamefont {Redondo}},\ and\ \bibinfo {author}
  {\bibfnamefont {J.}~\bibnamefont {Isern}},\ }\bibfield  {title} {\bibinfo
  {title} {{New axion and hidden photon constraints from a solar data global
  fit}},\ }\href {https://doi.org/10.1088/1475-7516/2015/10/015} {\bibfield
  {journal} {\bibinfo  {journal} {JCAP}\ }\textbf {\bibinfo {volume} {10}},\
  \bibinfo {pages} {015}},\ \Eprint {https://arxiv.org/abs/1501.01639}
  {arXiv:1501.01639 [astro-ph.SR]} \BibitemShut {NoStop}%
\bibitem [{\citenamefont {Chang}\ \emph {et~al.}(2017)\citenamefont {Chang},
  \citenamefont {Essig},\ and\ \citenamefont {McDermott}}]{Chang:2016ntp}%
  \BibitemOpen
  \bibfield  {author} {\bibinfo {author} {\bibfnamefont {J.~H.}\ \bibnamefont
  {Chang}}, \bibinfo {author} {\bibfnamefont {R.}~\bibnamefont {Essig}},\ and\
  \bibinfo {author} {\bibfnamefont {S.~D.}\ \bibnamefont {McDermott}},\
  }\bibfield  {title} {\bibinfo {title} {{Revisiting Supernova 1987A
  Constraints on Dark Photons}},\ }\href
  {https://doi.org/10.1007/JHEP01(2017)107} {\bibfield  {journal} {\bibinfo
  {journal} {JHEP}\ }\textbf {\bibinfo {volume} {01}},\ \bibinfo {pages}
  {107}},\ \Eprint {https://arxiv.org/abs/1611.03864} {arXiv:1611.03864
  [hep-ph]} \BibitemShut {NoStop}%
\bibitem [{\citenamefont {Hong}\ \emph {et~al.}(2021)\citenamefont {Hong},
  \citenamefont {Shin},\ and\ \citenamefont {Yun}}]{Hong:2020bxo}%
  \BibitemOpen
  \bibfield  {author} {\bibinfo {author} {\bibfnamefont {D.~K.}\ \bibnamefont
  {Hong}}, \bibinfo {author} {\bibfnamefont {C.~S.}\ \bibnamefont {Shin}},\
  and\ \bibinfo {author} {\bibfnamefont {S.}~\bibnamefont {Yun}},\ }\bibfield
  {title} {\bibinfo {title} {{Cooling of young neutron stars and dark gauge
  bosons}},\ }\href {https://doi.org/10.1103/PhysRevD.103.123031} {\bibfield
  {journal} {\bibinfo  {journal} {Phys. Rev. D}\ }\textbf {\bibinfo {volume}
  {103}},\ \bibinfo {pages} {123031} (\bibinfo {year} {2021})},\ \Eprint
  {https://arxiv.org/abs/2012.05427} {arXiv:2012.05427 [hep-ph]} \BibitemShut
  {NoStop}%
\bibitem [{\citenamefont {Schwarz}\ \emph {et~al.}(2015)\citenamefont
  {Schwarz}, \citenamefont {Knabbe}, \citenamefont {Lindner}, \citenamefont
  {Redondo}, \citenamefont {Ringwald}, \citenamefont {Schneide}, \citenamefont
  {Susol},\ and\ \citenamefont {Wiedemann}}]{Schwarz:2015lqa}%
  \BibitemOpen
  \bibfield  {author} {\bibinfo {author} {\bibfnamefont {M.}~\bibnamefont
  {Schwarz}}, \bibinfo {author} {\bibfnamefont {E.-A.}\ \bibnamefont {Knabbe}},
  \bibinfo {author} {\bibfnamefont {A.}~\bibnamefont {Lindner}}, \bibinfo
  {author} {\bibfnamefont {J.}~\bibnamefont {Redondo}}, \bibinfo {author}
  {\bibfnamefont {A.}~\bibnamefont {Ringwald}}, \bibinfo {author}
  {\bibfnamefont {M.}~\bibnamefont {Schneide}}, \bibinfo {author}
  {\bibfnamefont {J.}~\bibnamefont {Susol}},\ and\ \bibinfo {author}
  {\bibfnamefont {G.}~\bibnamefont {Wiedemann}},\ }\bibfield  {title} {\bibinfo
  {title} {{Results from the Solar Hidden Photon Search (SHIPS)}},\ }\href
  {https://doi.org/10.1088/1475-7516/2015/08/011} {\bibfield  {journal}
  {\bibinfo  {journal} {JCAP}\ }\textbf {\bibinfo {volume} {08}},\ \bibinfo
  {pages} {011}},\ \Eprint {https://arxiv.org/abs/1502.04490} {arXiv:1502.04490
  [hep-ph]} \BibitemShut {NoStop}%
\bibitem [{\citenamefont {An}\ \emph {et~al.}(2016)\citenamefont {An} \emph
  {et~al.}}]{JUNO:2015zny}%
  \BibitemOpen
  \bibfield  {author} {\bibinfo {author} {\bibfnamefont {F.}~\bibnamefont {An}}
  \emph {et~al.} (\bibinfo {collaboration} {JUNO}),\ }\bibfield  {title}
  {\bibinfo {title} {{Neutrino Physics with JUNO}},\ }\href
  {https://doi.org/10.1088/0954-3899/43/3/030401} {\bibfield  {journal}
  {\bibinfo  {journal} {J. Phys. G}\ }\textbf {\bibinfo {volume} {43}},\
  \bibinfo {pages} {030401} (\bibinfo {year} {2016})},\ \Eprint
  {https://arxiv.org/abs/1507.05613} {arXiv:1507.05613 [physics.ins-det]}
  \BibitemShut {NoStop}%
\bibitem [{\citenamefont {Kim}\ \emph {et~al.}(2016)\citenamefont {Kim} \emph
  {et~al.}}]{NEOS:2015dzs}%
  \BibitemOpen
  \bibfield  {author} {\bibinfo {author} {\bibfnamefont {B.~R.}\ \bibnamefont
  {Kim}} \emph {et~al.} (\bibinfo {collaboration} {NEOS}),\ }\bibfield  {title}
  {\bibinfo {title} {{Development and Mass Production of a Mixture of LAB- and
  DIN-based Gadolinium-loaded Liquid Scintillator for the NEOS Short-baseline
  Neutrino Experiment}},\ }\href {https://doi.org/10.1007/s10967-016-4826-1}
  {\bibfield  {journal} {\bibinfo  {journal} {J. Radioanal. Nucl. Chem.}\
  }\textbf {\bibinfo {volume} {310}},\ \bibinfo {pages} {311} (\bibinfo {year}
  {2016})},\ \Eprint {https://arxiv.org/abs/1511.05551} {arXiv:1511.05551
  [physics.ins-det]} \BibitemShut {NoStop}%
\end{thebibliography}%




\end{document}